\documentclass[fleqn]{2023SCGE}
\usepackage[utf8]{inputenc}
\usepackage{cite}
\hypersetup{hidelinks}
\begin{document}
\ArticleType{Article}
\title{Pseudo-critical localization}{Pseudo-critical localization}
\author[1]{Hui-Qiang Liang}{}
\author[2]{Linhu Li}{lilinhu@quantumsc.cn}
\author[1]{Guo-Fu Xu}{xgf@sdu.edu.cn}
\AuthorMark{Liang H Q}
\AuthorCitation{H. Q. Liang, L. Li, and G. F. Xu}
\address[1]{Department of Physics, Shandong University, Jinan 250100, China}
\address[2]{Quantum Science Center of Guangdong-Hong Kong-Macao Greater Bay Area (Guangdong), Shenzhen, China}
\abstract{We uncover a novel localization phenomenon, termed pseudo-critical localization, in non-Hermitian critical systems. This phenomenon is induced by the synergistic interplay between weak localization and non-Hermitian skin effects. It exhibits critical-like behavior at finite scales and undergoes a crossover to Anderson localization as the system size increases, which is in sharp contrast to critical localization that universally exhibits persistent criticality in the thermodynamic limit. Concretely, we use the non-Hermitian quasiperiodic mosaic model and non-reciprocal Aubry--Andr\'e model as minimal models to present the general features of size-dependent critical localization, and use entanglement entropy analysis to confirm the size-dependent nature. This discovery challenges the established paradigm of critical localization, opening new avenues for understanding and controlling critical behavior in non-Hermitian systems.}
\maketitle
\Authorfootnote
\begin{multicols}{2}

\section{Introduction}
Critical localization in Hermitian systems is of great interest in localization physics and has been attracting tremendous attention~\cite{anderson1958absence,kazushige1973localization,THOULESS197493,evers2008anderson}. Critical localization wavefunctions display multifractality, characterized by infinite sets of fractal dimensions~\cite{mandelbrot1974intermittent,mandelbrot1983fractal,falconer2013fractal}. They are accompanied by singular continuous spectra of zero Lebesgue measure, belonging to neither the absolutely continuous nor the pure-point type ~\cite{hastsugai1990energy,han1994critical,tanese2014fractal,liu2015localization,zeng2016generalized,
wang2016phase,yang2017dynamical,yao2019critical,wang2020realization,goblot2020emergence,xiao2021observation,
tong2021dynamics,liu2022anomalous,wang2022quantum,gon2023critical,zhou2023exact,li2023observation,
lin2023general,lee2023critical,dai2023emergence,shimasaki2024anomalous,li2024mapping,chen2024multifractality,
gou2024multiple,yang2024exploring,duncan2024critical,qi2024wave,bai2025tunably,dotti2025measuring}. Notably, these intriguing features are size-independent, that is, they persist robustly in the thermodynamic limit, as established across systems ranging from quasiperiodic models to disordered ones. This size-independent feature originates from the scale-invariant renormalization group flow near a fixed point, where critical wavefunctions (e.g., critical states at Anderson transition) exhibit thermodynamic-limit self-consistency due to the divergence of correlation length~\cite{fisher1974rg,RG2023QPS}.

While critical localization in Hermitian systems has been extensively studied, research has expanded to its counterparts in non-Hermitian systems. 
Non-Hermitian systems host many exotic features~\cite{lee2016anomalous,lee2019anatomy,yao2018edge,yokomizo2019non,jiang2019interplay,borgnia2020non,okuma2020topological,
ashida2020non,li_criticalNHSE_2020,zhang2020correspondence,zhang2021observation,delplace2021symmetry,mandal2021symmetry,zhang2022universal,
zhang2022review,lin2023topological,nakamura2024BBC,longhi2025erratic,liang2025intrinsic,ou2025ASL,yoshida2025hopf}, and are known for their extraordinary sensitivity to boundary conditions~\cite{budich2020nonH,mcdonald2020exponentially,koch2022quan,bao2022exp,yuan2023non,
nakai2024topo,bao2024exponential,maddi2024exact,parto2025enhanced} and system sizes~\cite{li_criticalNHSE_2020,gou2021exact,li2021impurity,yokomizo_scaling_2021,rafi-ul-islam_system_2022,qin_universal_2023,li2023scale,gou2023accumulation,
guo2024scale,Y2024from,liang2025intrinsic,ou2025ASL,zhou2026critical}, which fundamentally disrupt many paradigms established in Hermitian physics, such as the bulk-boundary correspondence in topological phases~\cite{yao2018edge,liang2025intrinsic} and critical phenomena~\cite{li_criticalNHSE_2020,zhou2026critical}. 
In non-Hermitian quasiperiodic or disordered systems, previous studies have revealed rich phenomena arising from the interplay between non-Hermiticity and quasiperiodicity/disorder, including phase coexistence~\cite{longhi2021phase,liu2021localization,liu2024emergent,sun2024hybrid,li2024ring,
gandhi2024nonhermitian,li2024asymmetric,shu2025multifractality,zheng2025emergent,zhao2025fate,dong2025critical} and mutual transformation of different states~\cite{jiang2019interplay,cai2022localization,longhi2025erratic,zhang2025scale}.
Despite these advances, established research on non-Hermitian systems has so far only identified critical localization with size-independent critical properties that persist robustly in the thermodynamic limit, or size-dependent behaviors~\cite{li_criticalNHSE_2020,gou2021exact,li2021impurity,yokomizo_scaling_2021,qin_universal_2023,
li2023scale,gou2023accumulation,guo2024scale,Y2024from,liang2025intrinsic,ou2025ASL} in the absence of critical localization marked by fractal characteristics of the states.
However, whether critically localized states within the critical region retain their Hermitian characteristics in the non-Hermitian context remains an open question. In particular, the scale-invariant nature of  critical localization seems to contradict the size-dependent behaviors induced by non-Hermiticity. In fact, we show that this is not necessarily the case: there exists a class of critical-like states in one-dimensional non-Hermitian quasiperiodic systems which, under many circumstances, can be mistaken for critical states. This discovery fundamentally renews our understanding of critical phenomena in non-Hermitian systems.

\end{multicols}
\begin{figure}[H]
\centering
\includegraphics[width=11cm]{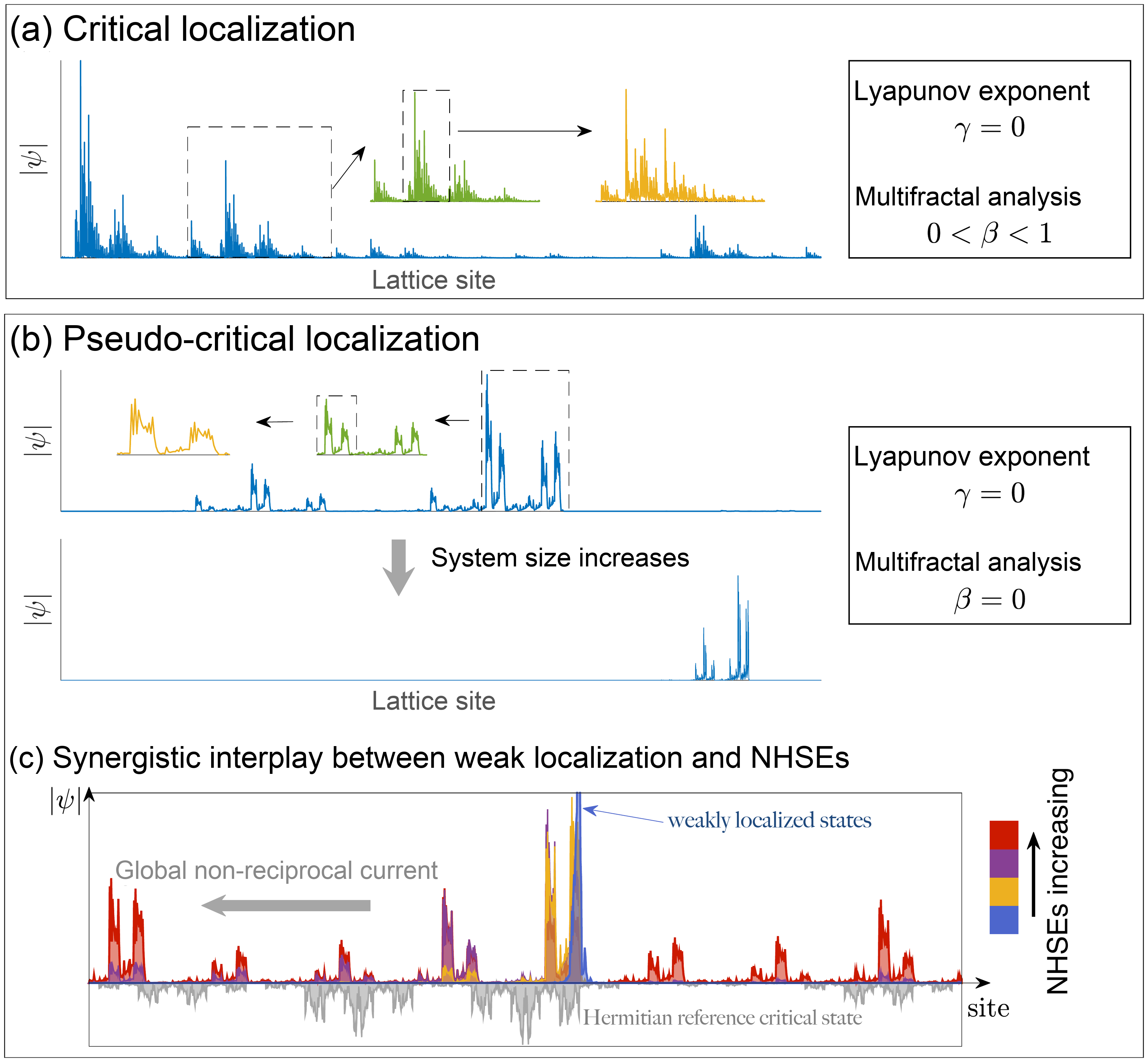}
\caption{(a) Critical localization: The wavefunction possesses self-similarity, characteristic of multifractality; the small profiles connected by arrows illustrate magnified portions of the wavefunction. (b) Pseudo-critical localization: The wavefunction exhibits critical-like behavior only at finite sizes, crossing over to a localized state in the thermodynamic limit. The upper small profiles illustrate finite-size self-similarity, while the lower profile shows localization at a larger size. {(c) The mechanism: Local non-reciprocal hopping induces weak localization, while the global non-reciprocal current induced by NHSEs extends its spatial range, giving a localization length comparable to the system size and producing pseudo-critical states. The gray profile below the horizontal axis shows the amplitude $|\psi_j|$ of a representative critical eigenstate of the Hermitian reference Hamiltonian $H_0$ ($h=g=0$), illustrating the underlying critical-state structure before non-Hermiticity is introduced. The profile is inverted only for visual clarity and does not indicate negative amplitudes.}
}
\label{fig1}
\end{figure}
\begin{multicols}{2}

Distinct from critical localization, whose features are schematically shown in Fig.~\ref{fig1}(a), this localization exhibits critical-like behavior only at finite sizes and crosses over to Anderson localization in the thermodynamic limit, as shown in Fig.~\ref{fig1}(b). Throughout this work, we use $L_c$ to denote the finite-size crossover scale separating the critical-like window from the asymptotic localized regime, rather than a critical length associated with a renormalization-group (RG) fixed point. Due to its size-dependent character and its ability to mimic critical states at finite sizes, we term this phenomenon pseudo-critical localization. Its physical origin lies in the interplay between weak localization and non-Hermitian skin effects (NHSEs) in critical systems, as sketched in Fig.~\ref{fig1}(c), where the skin-induced non-reciprocal current extends the spatial range of weakly localized states, giving them localization lengths comparable to the system size. We use the non-Hermitian quasiperiodic mosaic and non-reciprocal Aubry-André models to demonstrate the general features of pseudo-critical localization, with entanglement entropy confirming its size-dependent nature.

\section{Model and localization diagnostics}
\begin{figure}[H]
\centering
\includegraphics[width=13cm]{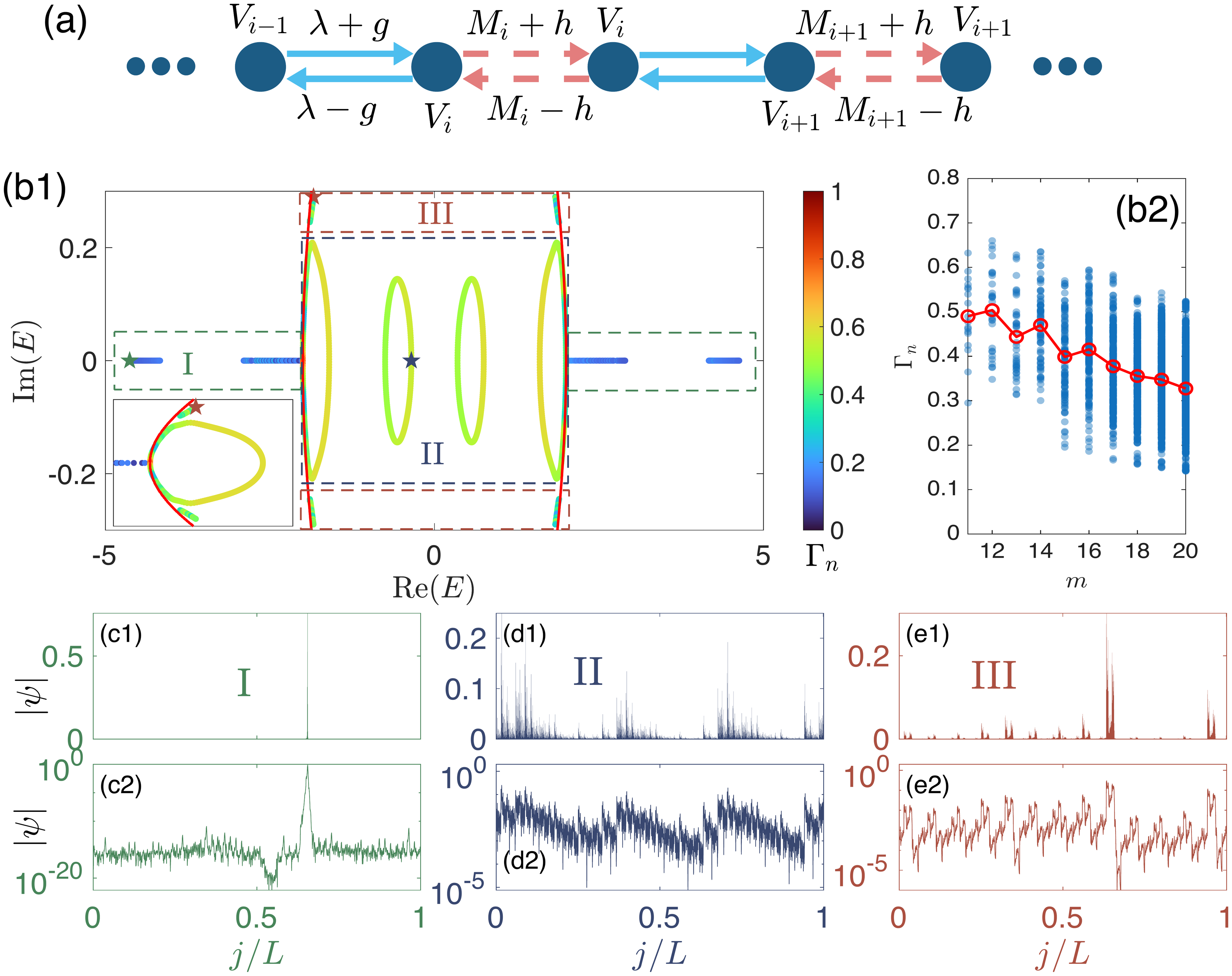}
\caption{(a) The non-Hermitian quasiperiodic mosaic model with $\kappa =2$. The solid circles denote the lattice sites, $V_i$ is the quasiperiodic mosaic potential, and the red dotted and blue arrows denote the quasiperiodic hopping ($M_i\pm h$) and constant hopping ($\lambda\pm g$) respectively, where $M_i=V_i=2t_0\cos(2\pi\kappa\alpha i)$ at the $i$-th quasicell. (b1) The eigenenergies of $H$ and the associated FDs $\Gamma_n$ ($\Gamma_n\rightarrow 0$ for localized states, $\Gamma_n\rightarrow 1$ for extended states, and intermediate values for critical states). (b2) Scaling of $\Gamma_n$ for type-III eigenstates as the system size increases, where $N=F_m$ and $L=2F_m$. The scatter plot shows the $\Gamma$ for all type-III eigenstates at each system size, while the line shows the mean value of $\Gamma$. (c1,c2), (d1,d2), and (e1,e2) show the characteristic spatial profiles of types I, II, and III eigenstates, respectively, on linear (top) and logarithmic (bottom) scales. The inset in (b1) enlarges the spectral region containing the representative eigenenergies marked by stars; their colors identify the states shown in (c)--(e). Other parameters are $\lambda = 2$, $h = 0.7$, $g = 0.73$, $\theta=0$, and $N=987$.}
\label{fig2}
\end{figure}

We now consider the non-Hermitian quasiperiodic mosaic model, as pictorially shown in Fig.~\ref{fig2}(a). The Hermitian part of the model, which hosts the transition between Anderson localization and critical localization, is described by~\cite{zhou2023exact}
\begin{equation}
H_0 = \sum_j\left(t_j\hat{a}_j^\dagger\hat{a}_{j+1}+\text{H.c.}\right)+\sum_jV_j\hat{a}_j^\dagger\hat{a}_j,
\label{eq:H0}
\end{equation}
where $\hat{a}_j^\dag$ $(\hat{a}_j)$ is the creation (annihilation) operator associated with the $j$-th site, with $j=1,\dots,L$. The quasiperiodic hopping coefficient $t_j$ and on-site potential $V_j$ are mosaic, with
\begin{equation}
\{t_j,V_j\} = \left\lbrace
\begin{aligned}
\ &\{\lambda,2t_0\cos(2\pi\alpha (j-1)+\theta)\},	&j = 1\ \text{mod}\ \kappa,\\
\ &2t_0\cos(2\pi\alpha j+\theta)\{1,1\},\quad &j = 0\ \text{mod}\ \kappa,\\
\ &\{\lambda,0\},\qquad &\text{else}.
\end{aligned}
\right.
\end{equation}
Here, $\lambda$ and $\theta$ denote the hopping coefficient and phase offset, respectively, and $t_0=1$ is set as the energy unit. 
$\kappa \geq 2$ is an integer representing how many lattice sites are within a quasicell, that is, $L=\kappa{N}$ with $N$ representing the number of quasicells. 
For simplicity, here we focus on the model with $\kappa=2$; extensions of our results to general values of $\kappa$ can be found in {Supplemental Section S1}.
The quasiperiodic modulation is described by an irrational number $\alpha$, 
chosen as 
$\alpha = \lim_{m\rightarrow\infty}(F_{m-1}/F_{m})=(\sqrt{5}-1)/2$, where $F_m$ are Fibonacci numbers defined recursively by $F_{m+1} = F_{m-1} + F_m$, starting from $F_0 = F_1 = 1$.
In finite systems, one may choose the number of quasicells  $N$ as $F_m$ and  take the rational approximation $\alpha \simeq \alpha_m = F_{m-1}/{F_m}$.

We introduce non-reciprocal hoppings to the mosaic model, and the total Hamiltonian is given by
\begin{equation}
\begin{aligned}
H = &H_0 + \sum_{j=0\ \text{mod}\ \kappa}\left( h \hat{a}_j^\dag\hat{a}_{j+1}-h\hat{a}_{j+1}^\dag\hat{a}_{j}\right)\\
&+\sum_{j\neq0\ \text{mod}\ \kappa}\left( g \hat{a}_j^\dag\hat{a}_{j+1}-g\hat{a}_{j+1}^\dag\hat{a}_{j}\right),
\end{aligned}
\label{eq:model}
\end{equation}
where $h\ (h<2t_0)$ and $g\ (g<\lambda)$ control the magnitudes of the non-reciprocity. 
As shown in Fig.~\ref{fig2}(b), the eigenenergies of $H$ can be divided into three categories (labeled I, II, and III) according to their spectral characteristics and the fractal dimension (FD) $\Gamma_n$ of the corresponding eigenstate, where $\Gamma_n = -\ln\big[ \sum_{j=1}^L|\psi_{j,n}|^4\big]\big/\ln L$ with $\psi_{j,n}$ the amplitude of the $n$-th eigenstate at the $j$-th lattice site. We accordingly designate the associated eigenstates as types I, II, and III eigenstates. 

\end{multicols}
\begin{multicols}{2}

As shown in Fig.~\ref{fig2}(c), type I eigenstates exhibit Anderson localization induced by the quasiperiodic modulation. Moreover, Fig.~\ref{fig2}(b1) shows that the strong localization (indicated by ${\rm FD}\approx0$) of these states blocks any possible non-reciprocity current even when the boundaries are closed, as reflected by their line-like spectrum enclosing no area~\cite{zhang2020correspondence}. In contrast, Fig.~\ref{fig2}(d) shows that type II eigenstates exhibit critical distributions across the whole lattice [as in Fig. 1(a)].
Since these states are distributed across the whole lattice, they experience a global non-reciprocal current that leads to a loop-like spectrum when the boundaries are closed~\cite{zhang2020correspondence}, as shown in Fig. \ref{fig2}(b1).
Note that these states transform into skin states with line-like spectrum when the boundaries are open, which cuts off the non-reciprocal current~[see {Supplemental Section S2}].
Most intriguingly, Fig.~\ref{fig2}(e) shows that type III eigenstates display fractal features and spatial distributions similar to those of type II eigenstates. However, Fig.~\ref{fig2}(b1) shows that their line-like spectrum indicates the absence of a global non-reciprocal current, suggesting an exotic mixture of types I and II scenarios. In Fig.~\ref{fig2}(b2), we calculate the $\Gamma_n$ for individual type III eigenstates, which shows a state-dependent decrease as $L$ increases, suggesting a crossover from critical-like behaviors toward localization.

\section{Eigenstate characteristics}
To characterize the criticality and verify the crossover of the eigenstates, we employ LE analysis and multifractal analysis. LE serves as a highly effective metric for quantifying eigenstate localization, calculable via transfer matrices. In our case, for an eigenstate with energy $E$, the corresponding LE reads
$
\gamma_\epsilon(E) = \lim_{N\rightarrow\infty}\frac{1}{2\pi N}\int\ln\Vert \mathcal{T}_{N,1}(\theta+i\epsilon)\Vert d\theta.
$
In the above, $\mathcal{T}_{N,1} = \prod_{i=1}^N\mathcal{T}_i=\mathcal{T}_N\mathcal{T}_{N-1}\cdots \mathcal{T}_2\mathcal{T}_1$, $N$ is the number of quasicells, $\Vert\cdot\Vert$ denotes the square root of the largest eigenvalue of the matrix, and $\epsilon$ is the imaginary part of the complexified $\theta$, where $\mathcal{T}_i = T_{j_0+\kappa}T_{j_0+\kappa-1}\cdots T_{j_0}$ is the transfer matrix of the $i$-th quasicell, with $T_j$ the transfer matrix of the $j$-th lattice site, i.e., $(\psi_{j+1},\psi_j)^\text{T} = T_j(\psi_{j},\psi_{j-1})^\text{T}$, and $j_0$ the first site of each quasicell. Using Avila's global theory~\cite{avila2015global}, the LE $\gamma_\epsilon(E)$ becomes $\epsilon$-independent, and reads~\cite{zhou2023exact} (see Supplemental Section S1 for the derivation)
\begin{equation}
\gamma(E) = {\rm max}\left\{ \frac{1}{2}\ln \left|\left|\frac{E}{\lambda+g}\right|+\sqrt{\left(\frac{E}{\lambda+g}\right)^2
-\frac{\lambda-g}{\lambda+g}}\right|,0\right\}.
\label{eq:LE}
\end{equation}
The mobility ring can be determined by the solution of 
\begin{equation}
\left|\left|\frac{E}{\lambda+g}\right|+\sqrt{\left(\frac{E}{\lambda+g}\right)^2
-\frac{\lambda-g}{\lambda+g}}\right|=1.
\label{eq:MR}
\end{equation}
The mobility ring is a loop centered at the origin in the complex energy plane, 
separating regions with $\gamma(E)>0$ and those with $\gamma(E)=0$, where $\gamma(E)>0$ means the corresponding eigenstates are localized, and $\gamma(E)=0$ indicates delocalized states
with infinite localization lengths, a characteristic of extended or critical states. As shown in Fig. \ref{fig2}(b), type I eigenstates fall outside the mobility ring and thereby their LEs are given by $\gamma(E)>0$, consistent with their localized distributions discussed previously. On the other hand, types II and III eigenstates are inside the mobility ring and have $\gamma(E)=0$, suggesting they are either extended or critical states.

We next apply multifractal analysis~\cite{liu2022anomalous} to unveil the distinction between types II and III eigenstates. For an eigenstate $\lvert \psi_n\rangle$, a scaling exponent $\beta_{j,n}$ can be extracted from the on-site probability
$
P_{j,n} = |\psi_{j,n}|^2 \sim \left(\frac{1}{F_m}\right)^{\beta_{j,n}},
$
where $\psi_{j,n}$ is the amplitude of $\lvert \psi_n\rangle$ at site $j$. According to multifractal analysis, 
when the eigenstate $\lvert \psi_n\rangle$ is extended, the maximum of $P_{j,n}$ over $j$ scales as $\max_j(P_{j,n})\sim 1/F_m$, and $\beta_{\text{min},n} \equiv\min_j(\beta_{j,n})=1$.
For a localized eigenstate $\lvert \psi_n\rangle$, $P_{j,n}$ peaks at very few sites and is nearly zero at the other sites, yielding $\max_j(P_{j,n})\sim\text{constant}$ and $\beta_{\text{min},n}=0$.
Finally, $\beta_{\text{min},n}$ for a critical  eigenstate $\lvert \psi_n\rangle$ falls within the interval $(0,1)$. To reduce fluctuations, we use the average scaling exponent, defined by
\begin{equation}
\beta_{\text{min}} = \frac{1}{L^\prime}\sum_n{\vphantom{\sum}}\beta_{\text{min},n},
\end{equation}
to characterize the three types of eigenstates, where $L'$ denotes the number of eigenstates of the same type (type I, II, or III), and the summation runs over these states.

\begin{figure}[H]
\centering
\begin{center}
\includegraphics[width=12cm]{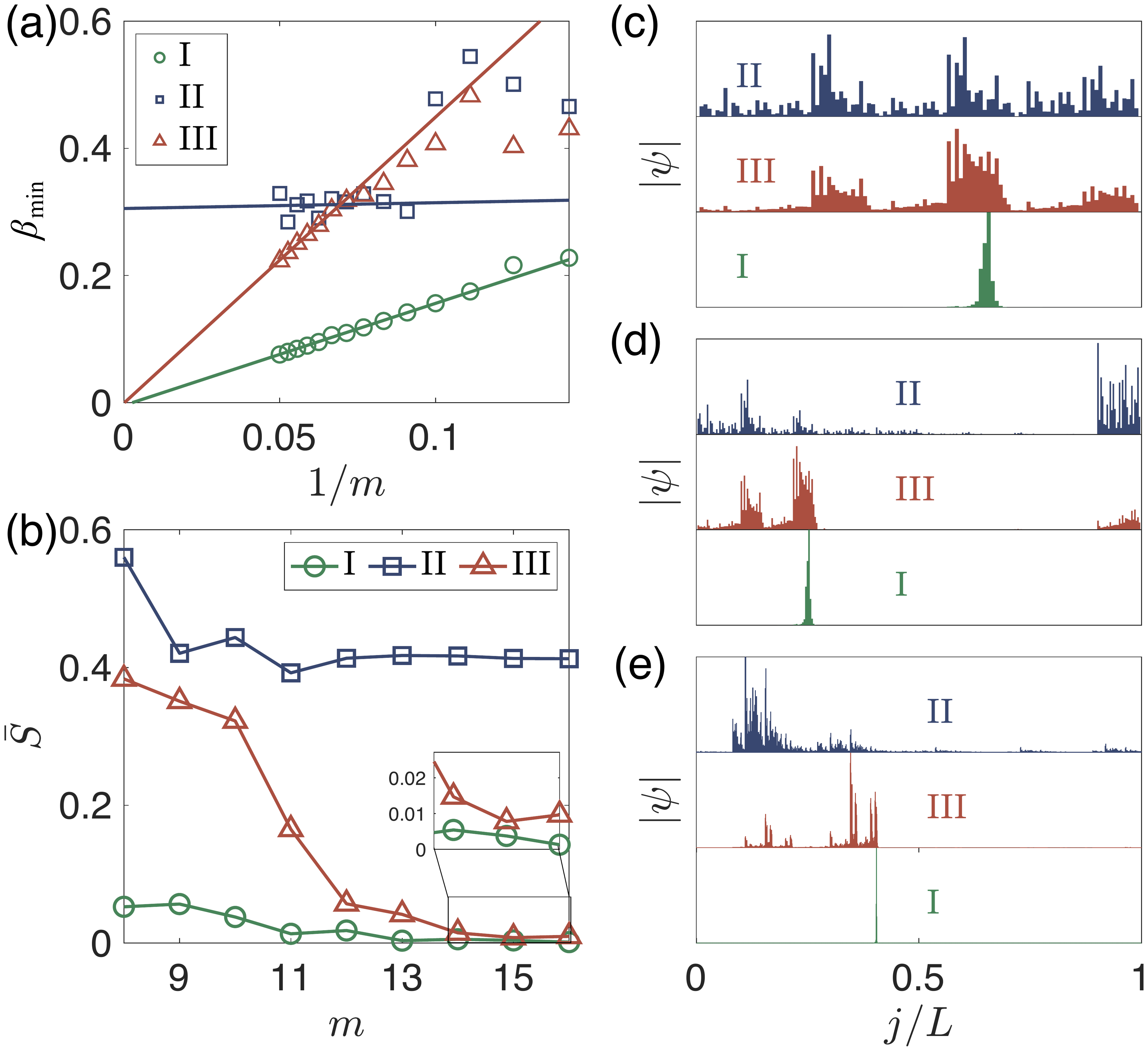}
\caption{(a) The average scaling exponents $\beta_{\rm min}$ as a function of the inverse Fibonacci index $1/m$ for the three types of eigenstates. Here the linear fit for type III eigenstates is obtained from the results with $m>13$ to show their asymptotic behavior in the thermodynamic limit; for smaller $m$, $\beta_{\rm min}$ for type III eigenstates deviates away from the linear fit and remains much less than $1$. The deviation at smaller $m$ occurs because these sizes are still within the finite-scale pseudo-critical window, where the effective multifractal properties are intrinsically size-dependent. This is quantitatively confirmed by our moments analysis in Supplemental Section S3. (b) Average entanglement entropy as a function of the Fibonacci index $m$. The inset enlarges the small-entropy region for types I and III at large $m$. (c-e) The distributions of the eigenstates with $m$ respectively being $9,11,16$. For (c-e), each comprises three panels: from top to bottom, they respectively exhibit the distribution of the conventional critical state (the $L/2$-th eigenstate), that of the pseudo-critical states (the eigenstate with the maximum imaginary energy), and that of the localized state (the first eigenstate). Here, we sort the eigenstates by the real parts of their energies. The parameters are the same as those in Fig.~\ref{fig2} except for $h = g = 0.75$.}
\label{fig3}
\end{center}
\end{figure}

Fig.~\ref{fig3}(a) shows the average scaling exponents $\beta_\text{min}$ for the three types of eigenstates, 
as a function of the inverse Fibonacci index $1/m$ that determines the system size through $L=\kappa F_m$ in our consideration. 
Consistent with the LE analysis, 
the average scaling exponents $\beta_\text{min}$ for types I and II eigenstates asymptotically tend to zero and have a finite value in the thermodynamic limit ($1/m\rightarrow 0$), which indicates their localized and critical natures, respectively. 
In comparison, $\beta_\text{min}$ of type III eigenstates asymptotically tends to zero when $1/m\rightarrow 0$, indicating their localized nature in the thermodynamic limit. $\beta_\text{min}$ of type III eigenstates can take values larger than that of type II eigenstates but much smaller than $1$ when $1/m\gtrsim0.07$, suggesting a critical-like behavior of these states at finite sizes. Indeed, our LE analysis shows that these states have $\gamma(E)=0$ and fall inside the mobility ring for a finite-size system, verifying their pseudo-critical behaviors.

\end{multicols}
\begin{multicols}{2}

\section{Size-dependent crossover}

To verify the size dependence of type III eigenstates, we analyze their biorthogonal entanglement entropy~\cite{chang2020entanglement,li_criticalNHSE_2020,li2022non,liang2022anomalous} as a function of system sizes. Specifically, we choose an entanglement partition between the left and right halves of the lattice (i.e., sites with $[1,L/2]$ and $[L/2+1,L]$), and the entanglement entropy is given by
\begin{equation}
S_n = -\sum_p \left[\zeta_{p,n}\ln\zeta_{p,n}+(1-\zeta_{p,n})\ln(1-\zeta_{p,n})\right],
\end{equation}
where $\zeta_{p,n}$ denotes the $p$-th eigenvalue of the correlator matrix $C_n$, whose $x,y$ entry is defined by
$
(C_{n})_{xy} = \langle \psi_n\rvert\hat{a}_x^\dagger\hat{a}_{y}\lvert \psi_n\rangle,
$
with $\lvert \psi_n\rangle$ the $n$-th eigenstate and $x,y\in[1,L/2]$.

Fig.~\ref{fig3}(b) shows the entanglement entropy averaged over each of the three types of eigenstates, $\bar{S}=\sum_{n}{\vphantom{\sum}}{S_n}/{L'}$, 
where type III eigenstates show a clear size-dependence in comparison with the others.
Explicitly, $\bar{S}\approx 0$ is seen for type I eigenstates, corresponding to the non-ergodic nature of localized states;
while type II eigenstates have a finite $\bar{S}$ smaller than its possible maximum $\ln 2$, corresponding to the partially ergodic nature of critical states with approximately extended distributions.
In contrast, $\bar{S}$ of type III eigenstates takes a finite value close to that of type II at small system sizes, but drops toward zero as the system size increases, showing a clear size-dependent crossover. We denote the corresponding crossover size by $L_c$ (in our numerics $m\simeq 12$), which should be viewed as a characteristic finite-size scale rather than a universal critical threshold.

In Figs.~\ref{fig3}(c) to (e), we display the wave amplitudes of the three types of eigenstates with different system sizes, where types I and II eigenstates show size-independent localized and critical distributions, respectively.
Meanwhile, type III eigenstates are seen to possess nonzero distribution across the system when $L<L_c$ [Figs.~\ref{fig3}(c) and (d)], indicating their partial ergodic and critical-like behaviors. However, as shown in Fig.~\ref{fig3}(e), wave amplitudes of type III eigenstate drastically drop to zero at most sites when $L>L_c$, indicating a non-ergodic localized feature.
This size-dependent localization crossover is unambiguously captured by the finite-size scaling of the average entanglement entropy, confirming that type III eigenstates are pseudo-critical states. {Further validations of the size-dependent crossover via the moments of pseudo-critical states and their spectral evolution as $L$ increases can be found in Supplemental Sections S3 and S4, respectively.}

\end{multicols}
\begin{figure}[H]
\begin{center}
\includegraphics[width=16.8cm]{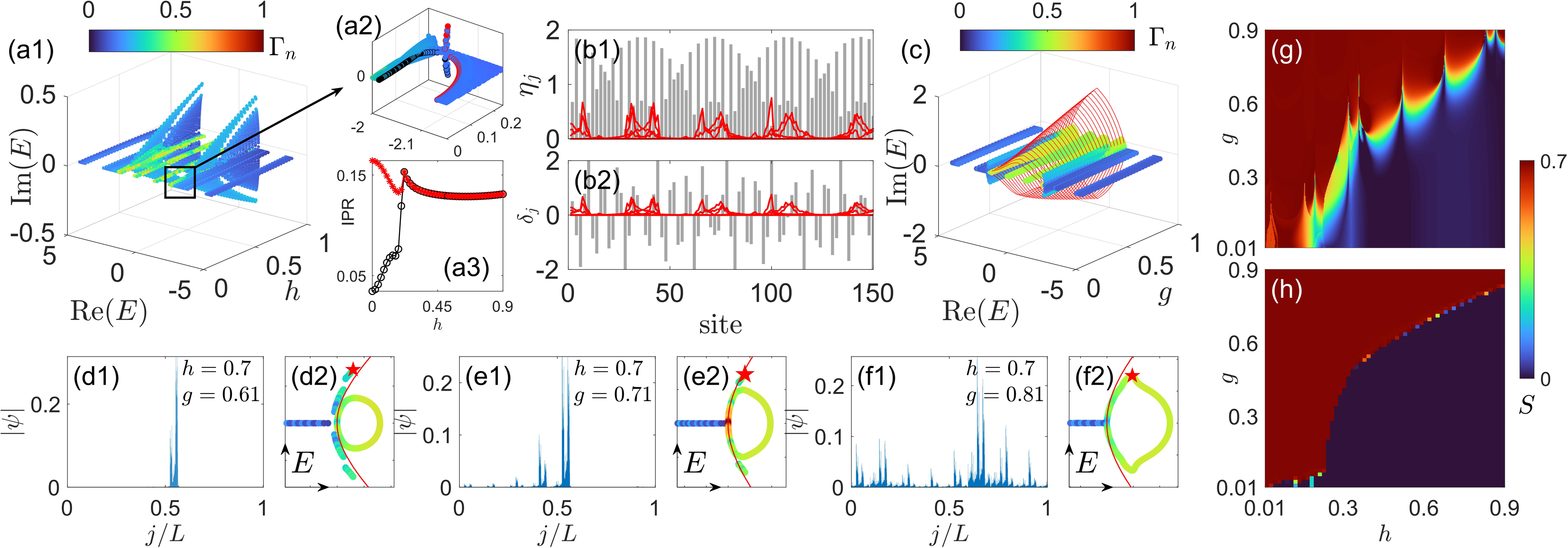}
\caption{(a1) FDs of different eigenstates as a function of $E$ and $h$ for $g = 0$ and $N = 1597$. (a2) Enlargement of the region boxed in (a1), showing the coalescence of eigenenergies. (a3) The inverse participation ratios, $\sum_j |\psi_{j,n}|^4$, of the eigenstates marked by red and black dots in (a2). (b1,b2) The average hopping amplitude $\eta_j$ and local non-reciprocity $\delta_j$, respectively, for $h = 0.7$, $g= 0$, and $j$ mod $\kappa=0$.
Red lines: the distributions of wavefunctions with $\left\vert \text{Re}(E)\right\vert >1.9$ and Im$(E) >0$. (c) FDs of different eigenstates as a function of $E$ and $g$ for $h = 0.7$ and $N = 1597$. Red loops: the MRs (Eq.~\eqref{eq:MR}). (d1)--(f1) Wavefunction distributions and (d2)--(f2) corresponding spectra with the MR (red curve); red stars identify the eigenenergies of the plotted states for $h=0.7$, $g=0.61,0.71,0.81$. (g) and (h) are $g$-$h$ phase diagrams of the eigenstates with max$($Im$(E))$ for sizes $m=9$ and $m=16$, respectively. Other parameters are $\lambda=2$, $\kappa=2$, and (d)--(f) $N=987$.}
\label{fig4}
\end{center}
\end{figure}
\begin{multicols}{2}

\section{Physical origin of pseudo-critical localization}

To elucidate the physical origin of pseudo-critical localization, we first examine the mosaic model without non-reciprocal hopping, governed by the Hamiltonian $H_0$. Previous research has established that this model hosts precise energy-dependent mobility edges separating localized and critical states, expressed as $|E|=\lambda$~\cite{zhou2023exact}. 
Further introducing non-reciprocity to hopping parameters may alter the distribution properties of eigenstates.
In our model, non-reciprocity is described by two parameters $h$ and $g$ in Eq.~\eqref{eq:model}.
To clarify the origin of pseudo-critical states, we first consider the case where $h\neq0$ and $g=0$, followed by the case where both $h$ and $g$ are nonzero.

As seen from Figs.~\ref{fig4}(a1) and (a2), adding a nonzero $h$ to the Hermitian Hamiltonian $H_0$ assigns complex eigenenergies to eigenstates of $H_0$ outside (but close to) the mobility edges $|E|=\lambda=2$. Furthermore, in the above process, one can see that the real eigenvalues coalesce at an exceptional point and subsequently split into complex-conjugate pairs with imaginary components of opposite signs, indicating the pseudo-Hermitian phase transition.
After this transition, critical and Anderson states (with real energies inside and outside the mobility edges, respectively) also coalesce into states with relatively weaker localization, as shown by the inverse participation ratios of two typical eigenstates in Fig. \ref{fig4}(a3).

To understand their origin, we define the local non-reciprocity $\delta_j$ and the average hopping amplitude $\eta_j$, 
\begin{align}
\delta_j = \ln\left|\frac{t_j+h}{t_j-h} \right|,\quad \eta_j=\sqrt{\left| (t_j+h)(t_j-h) \right|},
\end{align}
where $j\ \text{mod}\ \kappa=0$. 
Explicitly, the sign of $\delta_j$ describes the non-reciprocal direction (towards the left when $\delta_j> 0$), and its absolute value describes the net non-reciprocal strength; while $\eta_j$ describes the degree of coupling between two lattice sites connected by $t_j\pm h$. 
Note that besides the fully decoupled case with $(t_j, h)= 0$,
the condition $t_j=\pm h$ also indicates another effectively decoupled scenario,
where the two-site system is at an exceptional point and thus possesses a single eigenstate distributed only on one lattice site.
In Figs.~\ref{fig4}(b1) and (b2), it is seen that $\eta_j$ oscillates with a quasi-period of roughly $30\sim40$ lattice sites. Eigenstates with complex energies are found to localize around multiple positions where $\eta_j$ shows the strongest fluctuation. These localization centers coincide with DWs where $\delta_j$ undergoes sign reversal, indicating their origin in local non-reciprocal DWs.
Such a multi-center localization explains the weaker localization after the pseudo-Hermitian transition.

We now consider the case where $h\neq0$ and $g\neq0$. 
The parameter $g$ represents a uniform non-reciprocal pumping throughout the system, which can induce an NHSE for type II eigenstates under open boundary conditions. 
Physically, such a global non-reciprocal pumping tends to suppress the localization induced by the local hopping oscillation from $t_j$ and non-reciprocity from nonzero $h$. Indeed, increasing {$g$} enlarges the MR that encloses the critical region, as shown in Fig. \ref{fig4}(c).
Simultaneously, the eigenenergies of states weakly localized by local non-reciprocal DWs are gradually pushed towards the MR in the complex-energy plane. Once their eigenenergies enter the MR, these states evolve into pseudo-critical states within a finite-size window, as shown in Figs. \ref{fig4}(d) and (e).
Finally, further increasing $g$ merges these states into type II eigenstates with criticality, as shown in Fig. \ref{fig4}(f). 

{The crossover scale $L_c$ depends on the interplay between the local domain-wall confinement controlled by $h$ and the non-reciprocal pumping controlled by $g$. At fixed $h=0.75$, the wavefunction-moment results in {Supplemental Section S3} show that increasing $g$ shifts the crossover toward larger system sizes, consistent with an enhanced stretching of the weakly localized states. Conversely, reducing $g$ within the regime that retains a discernible pseudo-critical window can bring the crossover to smaller sizes. This provides a practical route to observing the crossover in finite experimental systems, where sizes on both sides of $L_c$ should be accessible.}

Figs.~\ref{fig4}(g) and (h) show the entanglement entropy $S_n$ of the eigenstate with the maximum imaginary energy at different sizes $m=9$ and $m=16$, respectively.
It is seen that pseudo-critical states appear around the crossover between regions with critical states ($S_n\approx \ln 2$) and Anderson states ($S_n\approx0$), indicated by the region with intermediate values of $S_n$ for a relatively smaller system size that disappears when the size increases to $m=16$.
We further note that the pseudo-critical localization is expected to be a general phenomenon in one-dimensional non-Hermitian quasiperiodic lattices which host critical localization as well as local non-reciprocal DWs and NHSEs. As an additional example, in {Supplemental Section S7}, pseudo-critical localization is shown to arise also in a non-reciprocal Aubry--Andr\'e model~\cite{li2024asymmetric}.

\section{Discussion and conclusion}

A novel localization phenomenon, termed pseudo-critical localization, has been identified in non-Hermitian critical systems. This phenomenon mimics critical localization only at finite sizes and undergoes a crossover to Anderson localization in the thermodynamic limit. Using the non-Hermitian mosaic and Aubry--Andr\'e models, we demonstrate that the underlying mechanism of pseudo-critical localization stems from the synergistic interplay between weak localization and NHSEs in critical systems. {Specifically, local non-reciprocal DWs induce eigenstates with inherent weak localization around these DWs. The global non-reciprocal current generated by the NHSE then extends the spatial range over which these states have appreciable amplitudes, giving them localization lengths comparable to the system size and transforming them into pseudo-critical states.}

{Recent experiments have demonstrated tunable exceptional points~\cite{liu2025tunableEP}, loss-induced hybrid-order skin effects~\cite{wu2025hybridorder}, and bound-pair phenomena driven by nonreciprocal two-body interactions~\cite{pu2026acousticpairs} in acoustic systems, as well as non-Hermitian Floquet topological sensing~\cite{zhou2025floquetsensors} and a non-Abelian Hatano--Nelson model~\cite{chen2026nonabeliancircuits} in electric circuits. These advances highlight the experimental control of non-Hermitian couplings and modes in classical platforms, motivating the exploration of pseudo-critical localization; a circuit implementation of our model is proposed in {Supplemental Section S5}.}

Beyond coexisting with quantum multifractality~\cite{castellani1989multifractal,ketz1992slow,ohtsuki1997anomalous,cuevas2001anomal,chab2008exper,
Rodriguez2009Multifractal,Faez2009obser,Lemari2009obser,Lemari2010critical,krav2010dynamical,Rodriguez2011multi,
muller2016critical,tarquini2017critical,ghosh2017coherent,akridas2019multi,max2023coherent,chen2023critical}, the pseudo-critical localization we discovered introduces a size-dependent crossover absent in finite-dimensional Hermitian systems. Strikingly, we note that critical properties without scale invariance have also been reported in Hermitian infinite-dimensional systems, where the system exhibits a logarithmic-size scaling in its eigenstate moments~\cite{chen2024quantum,chen2024describ}. In contrast, the pseudo-critical localization represents a crossover from critical-like states that show scale-invariant behaviors at small size to localized states at large sizes (see {Supplemental Section S3} for the scaling behaviors of eigenstate moments in our model). Furthermore,
these findings open avenues for wave manipulation in engineered systems and suggest the presence of hybrid skin-critical phenomena in higher dimensions where NHSEs become a universal feature and exhibit richer variations~\cite{zhang2021observation,zhang2022universal,zhang2022review,lin2023topological,zhang2025algebraic,liang2025intrinsic}. In interacting systems, the universal size-dependent features of this localization are expected to extend to interacting and many-body critical phases~\cite{luitz2015many,potter2015universal,bertrand2016anomalous,khemani2017critical,setiawan2017transport,kudo2018finite,
rispoli_quantum_2019,dumitrescu2019kt,mace2019multifractal,wang2020realization,wang2021many,xiao2021observation}, enriching the study of quantum thermalization in non-Hermitian systems~\cite{singha2025unveilingETH}. 

\Acknowledgements{ Guo-Fu Xu acknowledges support from Shandong Provincial Natural Science Foundation of China through Grant No. ZR2023MA048 and Shandong
Provincial Department of Science and Technology under
Project No. TQ012025002. Linhu Li acknowledges support from the
National Natural Science Foundation of China through Grant No. 12474159 and the Guangdong Provincial Quantum Science Strategic Initiative (Grant Nos. GDZX2504003 and GDZX2504006).}
\InterestConflict{The authors declare that they have no conflict of interest.}
\bibliographystyle{scpma}
\bibliography{refs}
\end{multicols}
\end{document}


\title{ Supplemental Materials for ``Pseudo-critical localization"}

\author{Hui-Qiang Liang}
\affiliation{Department of Physics, Shandong University, Jinan 250100, China}
\author{Linhu Li}\email{lilinhu@quantumsc.cn}
\affiliation{Quantum Science Center of Guangdong-Hong Kong-Macao Greater Bay Area (Guangdong), Shenzhen, China}
\author{Guo-Fu Xu}\email{xgf@sdu.edu.cn}
\affiliation{Department of Physics, Shandong University, Jinan 250100, China}

\renewcommand{\thefigure}{S\arabic{figure}}
\renewcommand{\theequation}{S\arabic{equation}}
\maketitle

\section*{S1. The calculation of the Lyapunov exponent}
In this section, we present the calculation of the Lyapunov exponent (LE) of the proposed model for an arbitrary value of $\kappa$. Consider the quasiperiodic Schrödinger operator for the model in the main text,
\begin{equation}
(H_{\lambda,\alpha,\theta}\psi)_n  = V_n\psi_n+t_n^+\psi_{n+1} + t_{n-1}^-\psi_{n-1},
\end{equation}
where
\begin{equation}
V_n(\theta) = \left\lbrace
\begin{aligned}
\ &2t_0\cos(2\pi\alpha (j-1)+\theta)	&j = 1\ \text{mod}\ \kappa,\\
\ &2t_0\cos(2\pi\alpha j+\theta) \quad &j = 0\ \text{mod}\ \kappa,\\
\ &0,\qquad &\text{else},
\end{aligned}
\right.
\end{equation}
and
\begin{equation}
t_n^\pm(\theta) = \left\lbrace
\begin{aligned}
\ &2t_0\cos(2\pi\alpha j+\theta) \pm h &j = 0\ \text{mod}\ \kappa,\\
\ &\lambda\pm g \quad &\text{else},
\end{aligned}
\right.
\end{equation}
with $\theta \in(0,2\pi]$, $\lambda\neq 0$ and $h\leq \lambda$. The LE can be calculated through the transfer matrices~\cite{zhou2023exact}, that is,
\begin{align}
&\gamma_\epsilon(E) = \lim_{L\rightarrow\infty}\frac{1}{2\pi N}\int\ln\Vert \mathcal{T}_{N,1}(\theta+i\epsilon)\Vert d\theta,\\
&\mathcal{T}_{N,1} = \prod_{i=1}^N\mathcal{T}_i=\mathcal{T}_N\mathcal{T}_{N-1}\cdots \mathcal{T}_2\mathcal{T}_1,~~\mathcal{T}_i = T_{j_0+\kappa}T_{j_0+\kappa-1}\cdots T_{j_0},~~T_j=
\begin{pmatrix}
\frac{E-V_j}{t_j^+}& \frac{t_{j-1}^-}{t_{j+1}^+}\\
1&0
\end{pmatrix}.
\nonumber
\end{align}
In the above, $\mathcal{T}_{N,1} = \prod_{i=1}^N\mathcal{T}_i=\mathcal{T}_N\mathcal{T}_{N-1}\cdots \mathcal{T}_2\mathcal{T}_1$, $N$ is the number of quasicells, $\Vert\cdot\Vert$ denotes the square root of the largest eigenvalue of the matrix, and $\epsilon$ is the imaginary part of the complexified $\theta$, where $\mathcal{T}_i = T_{j_0+\kappa}T_{j_0+\kappa-1}\cdots T_{j_0}$ is the transfer matrix of the $i$-th quasicell, with $T_j$ the transfer matrix of the $j$-th lattice site, i.e., $(\psi_{j+1},\psi_j)^\text{T} = T_j(\psi_{j},\psi_{j-1})^\text{T}$, and $j_0$ the first site of per quasicell. In the thermodynamic limit $L\rightarrow \infty$, the extended and critical states have $\gamma= 0$, while the localized states have $\gamma>0$. The transfer matrix $\mathcal{T}_i$ can be written as
\begin{equation}
\begin{aligned}
\mathcal{T}_i = \frac{\mathcal{T}_i^\prime}{(M+h)(\lambda+g)} = &\frac{\begin{pmatrix}
a_\kappa 	&\alpha a_{\kappa-1}\\
a_{\kappa-1}	&\alpha a_{\kappa-2}\\
\end{pmatrix}}{(M+h)(\lambda+g)}\\
&\times
\begin{pmatrix}
(E-M)^2-(M-h)(M+h)	&-(E-M)(\lambda-g)\\
(E-M)(\lambda+g)	&(\lambda-h)(\lambda+g)\\
\end{pmatrix},
\end{aligned}
\end{equation}
where $a_\kappa = \Delta^{-1}[((\beta+\Delta)/2)^{\kappa-1}-((\beta-\Delta)/2)^{\kappa-1}]$, $\Delta = \sqrt{\beta^2+4\alpha}$, $\beta = E/(\lambda+g)$, $\alpha=-(\lambda-g)/(\lambda+g)$ and $M(\theta+i\epsilon) = 2\cos (\theta+i\epsilon)$. Furthermore, we employ Avila’s global theory
of one-frequency analytical $SL(2,\mathbb{C})$ cocycle~\cite{avila2015global}. In the limit $\epsilon\rightarrow\infty$, a direct computation yields
\begin{equation}
\mathcal{T}^\prime_{i,\epsilon\rightarrow\infty}  = \begin{pmatrix}
a_\kappa 	&\alpha a_{\kappa-1}\\
a_{\kappa-1}	&\alpha a_{\kappa-2}\\
\end{pmatrix}
\begin{pmatrix}
-2\beta	&-\alpha	\\
-1		&0\\
\end{pmatrix}+\mathcal{O}(e^{-\epsilon}).
\end{equation}
Consequently, we have
\begin{equation}
\begin{aligned}
\kappa \gamma_{\epsilon\rightarrow\infty}(E) &= \ln \left\Vert \mathcal{T}^\prime_{i,\epsilon\rightarrow\infty}\right\Vert - A + \mathcal{O}(\epsilon^{-1})\\
&= \ln\ \left||a_{\kappa+1}|+\sqrt{a_{\kappa+1}^2-(-\alpha)^{\kappa-1}}\right| - A+ \mathcal{O}(\epsilon^{-1}).
\end{aligned}
\end{equation}
with 
\begin{equation}
A
=\frac{1}{2\pi}\int_0^{2\pi}\ln|(h + 2\cos(\theta + i\epsilon))|d\theta
= \left\lbrace \begin{matrix}
\ln{\frac{h+\sqrt{h^2-4}}{2}}, &h> 2,\\
0,	&h\leq2.\\
\end{matrix} \right.
\end{equation}

In the main text, we consider the case $h<2t_0$. According to Avila’s global theory, $\gamma_\epsilon(E)$ is a convex,  piecewise-linear function with respect to $\epsilon$. For energies $E$ within the spectrum, it follows that 
\begin{equation}
\kappa \gamma(E) = \max\left\{\ln\ \left||a_{\kappa+1}|+\sqrt{a_{\kappa+1}^2-(-\alpha)^{\kappa-1}}\right|,0\right\},
\end{equation}
with $\epsilon=0$. Since the localization length is given by $\xi = 1/\gamma$, a vanishing LE ($\gamma = 0$), which characterizes extended or critical states, corresponds to an infinite localization length ($\xi \to \infty$). Consequently, the mobility ring is determined by
\begin{equation}
\left||a_{\kappa+1}|+\sqrt{a_{\kappa+1}^2-(-\alpha)^{\kappa-1}}\right|=1,
\end{equation}
where $E =E_R + iE_I$, with $E_R$ and $E_I$ representing the real and imaginary parts of the eigenvalue, respectively. For $\kappa = 2$, the mobility ring is given by
\begin{equation}
\left|\left|\frac{E}{\lambda+g}\right|+\sqrt{\left(\frac{E}{\lambda+g}\right)^2-\frac{\lambda-g}{\lambda+g}}\right|=1,
\end{equation}
that is, the mobility ring is a closed loop in the complex energy plane, centered at $(E_R,E_I) = (0,0)$. The eigenstates associated with energies inside the loop exhibit a LE $\gamma = 0$, corresponding to extended or critical states, whereas those outside the loop have $\gamma > 0$, indicating localized behaviors.

\begin{figure}[htb]
\begin{center}
\includegraphics[width=16.8cm]{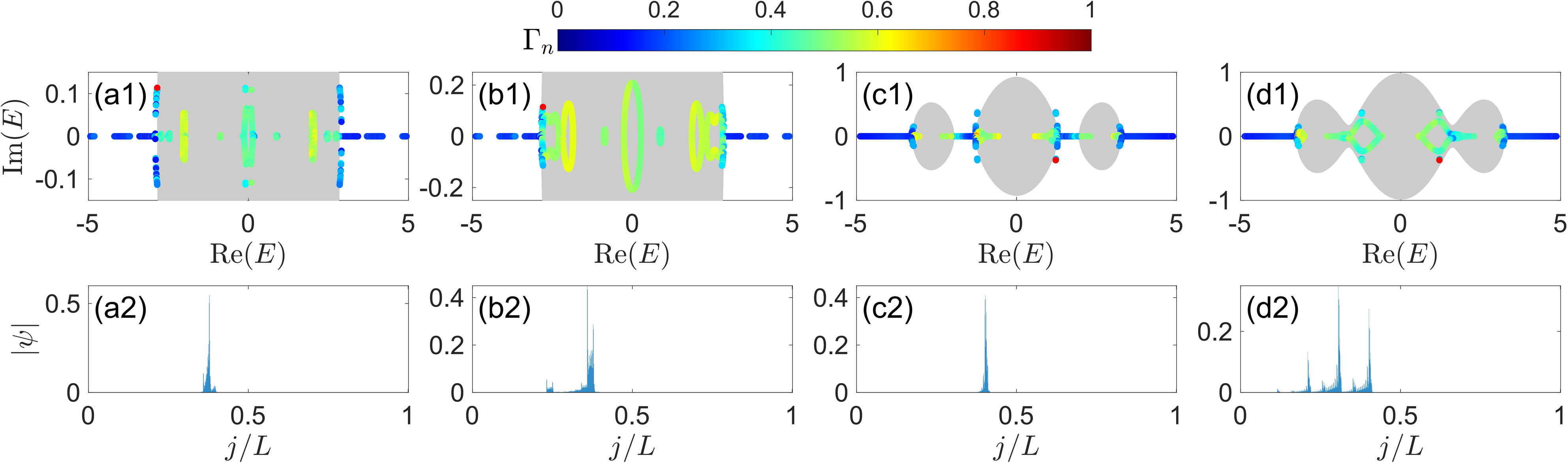}
\caption{Fractal dimensions $\Gamma_n$ of the eigenstates as a function of the complex energy $E$ and the corresponding mobility ring for $\kappa = 3$ (a,b) and $\kappa = 4$ (c,d). Gray regions in the complex energy plane indicate where the LE vanishes ($\gamma = 0$), with their boundaries forming the mobility ring. Wavefunction distributions corresponding to the energies marked in red in (a1, b1, c1, d1) are displayed in (a2, b2, c2, d2), respectively. Parameter values are: (a) $g = 0.1$; (b) $g = 0.4$; (c) $g = 0.1$; (d) $g = 0.3$. Other parameters are fixed at $N = 377$, $\lambda = 2$, and $h = 0.7$.}
\label{fig:highKappa}
\end{center}
\end{figure}

In Fig.~\ref{fig:highKappa}, we present numerical results for $\kappa = 3$ and $\kappa = 4$. The eigenstates corresponding to energies inside the mobility ring exhibit multifractal behaviors, as shown in Figs.~\ref{fig:highKappa}(b2) and (d2). Both cases are consistent with the analysis and argument presented in the main text.

\section*{S2. Pseudo-critical states under open boundary conditions}
In the main text, we studied pseudo-critical states under periodic boundary conditions (PBCs). In this section, we extend our analysis to the spectrum and localization behavior of pseudo-critical states under open boundary conditions (OBCs). We adopt the same Hamiltonian parameters as in Eq.~(3) of the main text and impose OBCs for this investigation.

\begin{figure}[htb]
\begin{center}
\includegraphics[width=16.8cm]{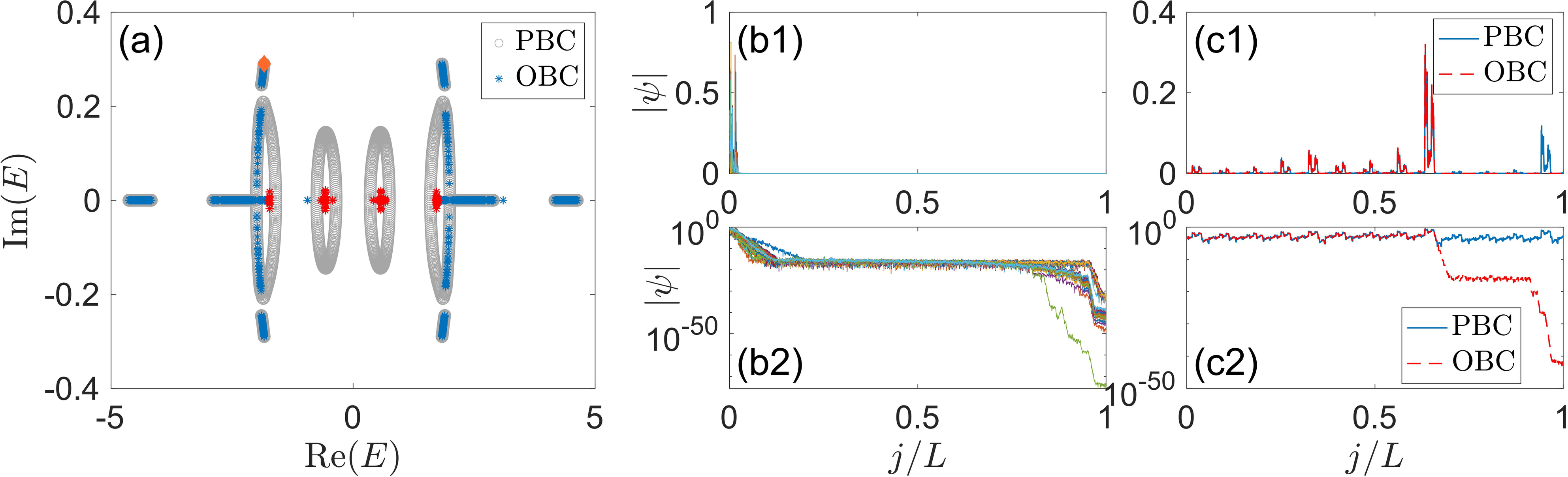}
\caption{(a) The spectra under PBCs and OBCs are marked by gray circles and blue asterisks, respectively. The eigenenergies of critical states under OBCs are marked by red asterisks, and the eigenenergies of the pseudo-critical states are marked by orange diamonds. (b) The wavefunction distributions of the critical states under OBCs. (c) The wavefunction distributions of the pseudo-critical states (their eigenenergies are marked by orange diamonds in (a)). Panels (b1,c1) use linear vertical scales, while (b2,c2) use logarithmic vertical scales. The other parameters are the same as those in Fig.~2 of the main text.}
\label{fig:OBC}
\end{center}
\end{figure}

As shown in Figs.~\ref{fig:OBC}(a) and (b), the critical states transform into skin states under OBCs, and their spectrum no longer retains a loop-like configuration. In contrast, as shown in Figs.~\ref{fig:OBC}(a) and (c), the spectrum of the pseudo-critical states remains largely unchanged under OBCs. The localization behavior of these pseudo-critical states under OBCs can be attributed to the presence of a domain wall in the bulk, which hinders the directional accumulation and reduces the stacking distance.

\begin{figure}[htb]
\begin{center}
\includegraphics[width=16.8cm]{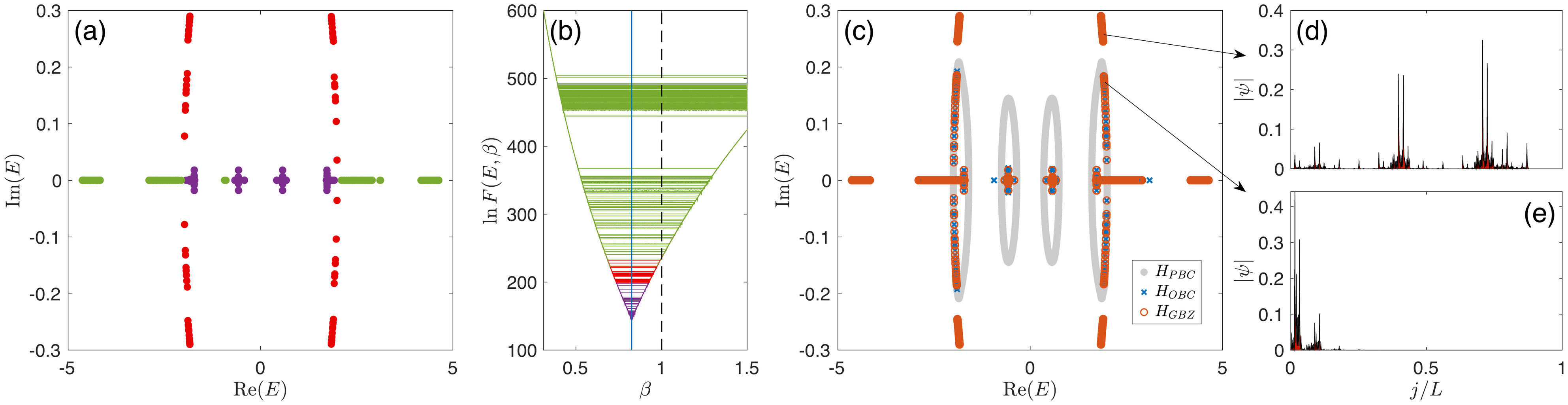}
\caption{(a) The eigenspectrum of $H_\text{PBC}$. The eigenspectrum of localized, weakly localized, and critical states are marked by green, red, and purple dots, respectively. (b) The evolution of $\ln F(E,\beta)$ with the increase of $\beta$. Here, $E$ is the eigenvalue of $H_\text{OBC}$. The light blue line is $\beta_\text{GBZ} \approx 0.8256$, which is the unified $\beta$ of the model. The black dash is $\beta = 1$, the Brillouin zone.
(c) The eigenspectrum of the model under different cases. (d) The distribution of wavefunction of pseudo-critical states with the model under OBCs. For the weakly localized states captured by loop spectrum under PBCs, this states (e) exhibit boundary accumulation when the model is under OBCs.
The other parameters are the same as those in Fig.~2 in the main text.}
\label{fig:GBZ}
\end{center}
\end{figure}

In non-Hermitian disordered systems, the localization length of skin modes can be captured by generalized Brillouin zone (GBZ) $\beta$. Here, we calculate $\beta$ for the model in the main text by modified GBZ theory~\cite{zhang2023bulk,liu2023mgbz}. Following the modified GBZ theory, we adopt the
transformation $t^\pm \beta^\pm$, where $\beta$ is the transformation coefficient and gives a quantitative description of the NHSE. One can get that the transformed Hamiltonian under PBC $[H_\text{GBZ}\equiv H_\text{PBC}(\beta)]$ satisfies
\begin{equation}
\det[E_\text{OBC}-H_\text{GBZ}] = \det[E_\text{OBC}-H_\text{OBC}] + f_\text{PBC},
\end{equation}
where $H_\text{PBC}(\beta) = S H_\text{PBC} S^{-1}$ with $S = \text{diag}[\beta,\beta^2,\cdots,\beta^N]$, and $f_\text{PBC}$ is an coefficient to be optimized by $\beta$. We mark $F(E_\text{OBC},\beta)=\left\vert \det[E_\text{OBC}-H_\text{PBC}(\beta)] - \det[E_\text{OBC}-H_\text{OBC}]\right\vert \equiv \left\vert f_\text{PBC}\right\vert$. For different $E_\text{OBC}$, the coefficient $\beta = \beta_\text{min}$ is available by minimizing $F(E_\text{OBC},\beta)$. The $E$-dependent of $\beta$ can be eliminated when a unified $\left\vert\beta\right\vert$ exists.
As shown in Figs.~\ref{fig:GBZ}(a) and (b), the global $\beta$ of this model is $\beta \sim 0.8256$ under the global minimum $\ln F(E,\beta)$. In Fig.~\ref{fig:GBZ}(c), one can see that the eigenenergies of the GBZ Hamiltonian almost perfectly match those of the OBC Hamiltonian.

\section*{S3. The scaling behavior of the moments of pseudo-critical states}

In a quasiperiodic lattice under periodic boundary conditions, reliable rational approximants of the irrational modulation require choosing system sizes along the Fibonacci sequence. Using non-Fibonacci sizes under PBCs leads to an uncontrolled phase mismatch at the boundary and can introduce spurious artifacts in eigenstates and multifractal observables. Therefore, in the below, we restrict $N$ to Fibonacci approximants to ensure numerical accuracy.

To obtain robust and smooth scaling trends without sacrificing the quasiperiodic approximation, we perform phase averaging: for each parameter set $(g,N)$, we draw phase parameter $\theta \in(0,2\pi]$ (a global phase shift of the quasiperiodic potential in the model) and average the relevant observables, eigenstate moments $P_q$ and fractal dimensions $D_q$. Note that in the main text we consider $\Gamma_n=D_2$. This $\theta$-averaging significantly suppresses sample-specific fluctuations and yields a stable estimation of the data.

\begin{figure}[htb]
\begin{center}
\includegraphics[width=7.8cm]{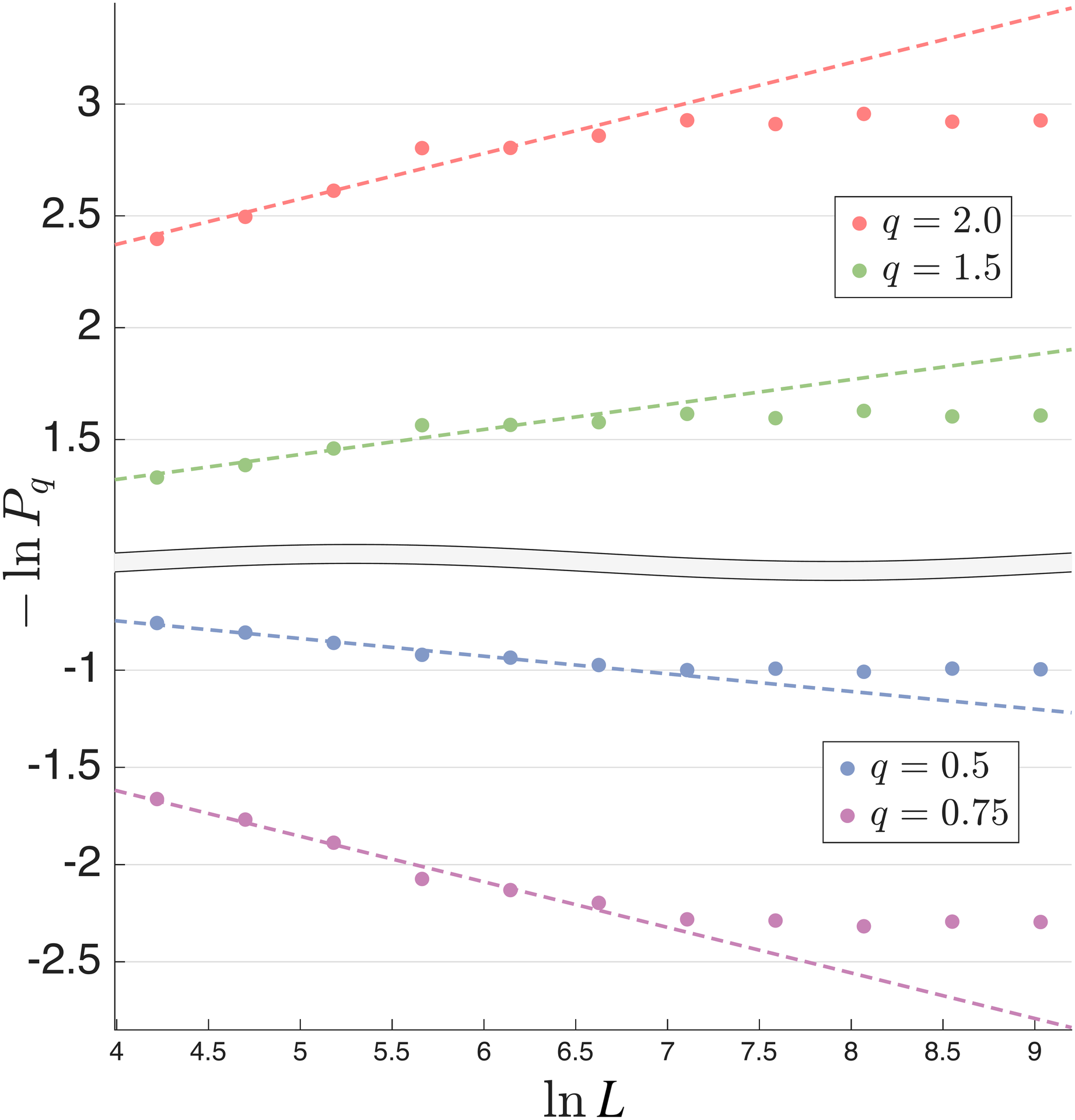}
\caption{Multifractality for moments of pseudo-critical states with $q=0.5,0.75,1.5,2.0$. The data are averaged over $500$ random phases $\theta\in(0,2\pi]$ for each size $L=2N$. The other parameters are the same as those in Fig.~3 in the main text.}
\label{fig:Pq}
\end{center}
\end{figure}

Fractal dimensions can be investigated by considering the moments $P_q \equiv\sum_j|\psi_j|^{2q}$ of order $q$ of on-site eigenstate amplitudes $|\psi_j|^2$ exhibiting an algebraic scaling behavior $P_q\sim L^{-D_q(q-1)}$. While $D_q=1$ for ergodic delocalization and $D_q = 0$ for localization, the multifractal dimension $D_q$ is a nontrivial function of $q$ with $0<D_q<1$ at the critical point. Fig.~\ref{fig:Pq} obviously show that the $D_q$ of pseudo-critical states are size-dependent. One can see that $D_q\approx 0$ when $\ln L > 7$. Note that the scaling behavior of pseudo‑critical states differs from the log‑multifractal properties exhibited by critical states in infinite‑dimensional systems. The latter is manifested as $P_q \sim (\ln L)^{-D_q(q-1)}$~\cite{chen2024quantum,chen2024describ}.
In the following, we focus on a single-parameter cut by fixing $h$ and varying $g$. This choice is intentional: the parameter $h$ controls the formation/strength of the local non-reciprocal domain-wall structure that induces weak localization in our mechanism, while $g$ primarily controls the strength of the NHSE. By fixing $h$ in a regime where the domain-wall–induced weak localization is already present, the $g$-sweep cleanly isolates the role of NHSE as a ``stretching" mechanism that enlarges the finite-size pseudo-critical window.

\begin{figure}[htb]
\begin{center}
\includegraphics[width=16.8cm]{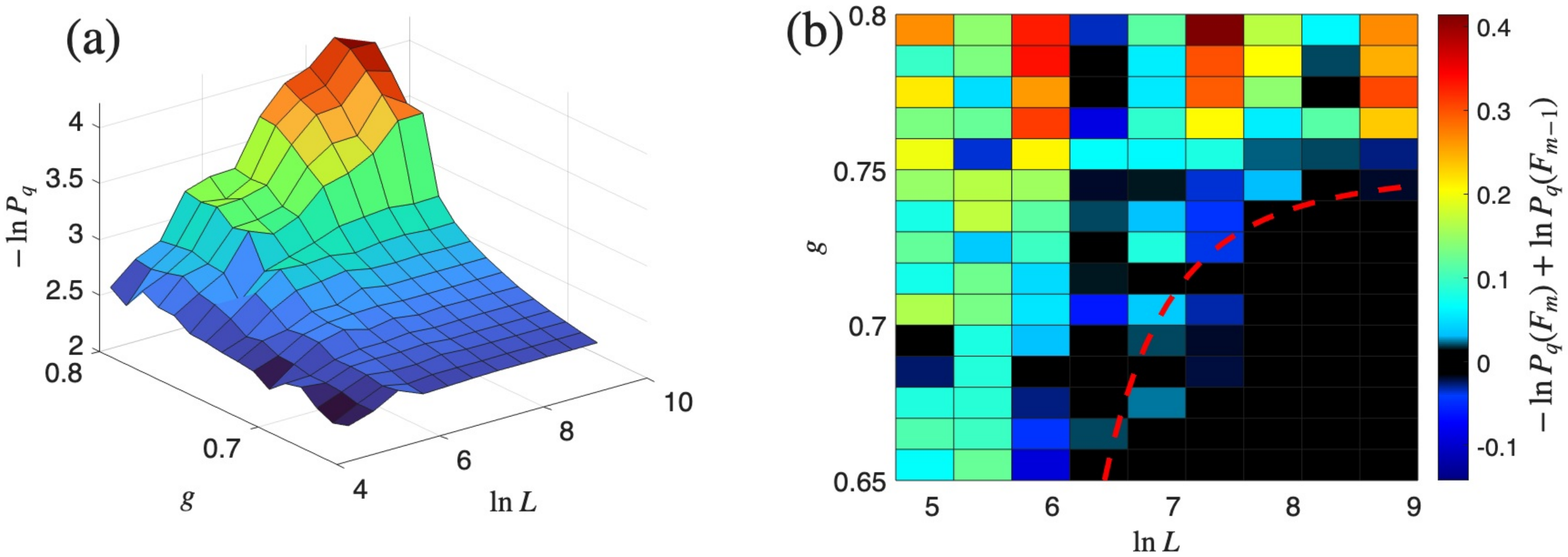}
\caption{Heatmap of the wavefunction moment for the pseudo-critical eigenstates at fixed $h=0.75$. (a) The color encodes $-\ln P_q$ (after averaging over 500 random quasiperiodic phases $\phi\in(0,2\pi]$) plotted as a function of the control parameter $g$ and the system size shown as $\ln L$ (with $L=\kappa N$ restricted to Fibonacci approximants). (b) The colorbar encodes the discrete logarithmic increment of the wavefunction moment between two consecutive Fibonacci system sizes $F_{m-1}$ and $F_m$. This quantity serves as an effective finite-size scaling indicator: values close to zero (dark region) indicate that the moment changes weakly between successive sizes and signal the onset of localization, while finite positive values indicate a stronger size dependence associated with critical-like scaling in a finite-size window. The red dashed curve in (b) indicates the approximate crossover front $g_c(L)$ estimated from the moment-based map and shown as a guide to the eye, e.g., $g_c(L) = g_\infty-A[\ln L-\ln L_0]^{-p} $ where $q=2$,$g_\infty=0.75$, $A=1$, $\ln L_0 = 4.5$ and $p=3.5$.}
\label{fig:S_L}
\end{center}
\end{figure}

Figure~\ref{fig:S_L}(a) shows that the wavefunction moments visualize the mechanism quantitatively: for small \(g\), the quantity \(-\ln P_q\) rapidly increases with system size \(L\), indicating that \(P_q\) quickly approaches its localized scaling behavior and thus the pseudo-critical window is small. As $g$ increases at the same fixed $h$, the moment-based map suggests that the state localization is induced at large $\ln L$, consistent with an increasing crossover scale $L_c$. This trend is consistent with our physical picture that enhancing NHSE at fixed domain-wall background promotes pseudo-critical localization by amplifying the macroscopic length scale, over which weak localization appears critical-like. For sufficiently large \(g\), the crossover length scale \(L_c\) is pushed toward the thermodynamic limit, suggesting that the pseudo-critical window no longer terminates at finite system sizes. In this regime, the system effectively exhibits scale-invariant behavior, consistent with a critical-like phase in the thermodynamic limit.

To further quantify the crossover discussed in Fig.~\ref{fig:S_L}(a), Fig.~\ref{fig:S_L}(b) plots the quantity
\[
-\ln P_q(F_m)+\ln P_q(F_{m-1}),
\]
namely the discrete logarithmic increment of the wavefunction moment $P_q$ between two consecutive Fibonacci system sizes. This quantity measures how sensitively the moments change with increasing system size, and therefore serves as an indicator of finite-size scaling. In particular, values close to zero indicate that the moments have become nearly size independent between successive Fibonacci approximants, signaling the onset of localization, whereas finite positive values indicate a pronounced finite multifractal dimension characteristic of the pseudo-critical regime.

From Fig.~\ref{fig:S_L}(b), one can clearly observe a crossover from a nearly zero region to a finite-valued region in the $(g,\ln L)$ plane. This crossover marks the boundary between the asymptoticly localized regime and the critical-like regime. In other words, the front \(g_c(L)\) approximately identifies the crossover from a localized regime, where the moments vary only weakly with size, to a pseudo-critical regime, where their finite-size scaling remains pronounced.

The tendency of $g_c(L)$ to approach a finite limiting value $g_\infty$ as $L$ increases is motivated by the spectral evolution shown in Fig.~4 in the main text. As $g$ increases, the loop spectrum expands and progressively absorbs the spectral sector associated with pseudo-critical localization; at the same time, the corresponding spectrum evolves from a line-like structure toward a loop-like one. This suggests that, in the large-$L$ limit, the crossover front should not drift indefinitely, but instead saturate toward a finite asymptotic value $g_\infty$.

Based on this physical picture, we adopt
\[
g_c(L)=g_\infty-A\,[\ln L-\ln L_0]^{-p}
\]
as an empirical ansatz,
which assumes that the difference \(g_\infty-g_c(L)\) decays algebraically with \(\ln L\). We emphasize that this form is not derived analytically, but is adopted as an empirical ansatz to summarize the crossover trend suggested by Fig.~\ref{fig:S_L}(b).

{\section*{S4. Size-dependent spectral evolution}}

\begin{figure}[htb]
\centering
\includegraphics[width=16.8cm]{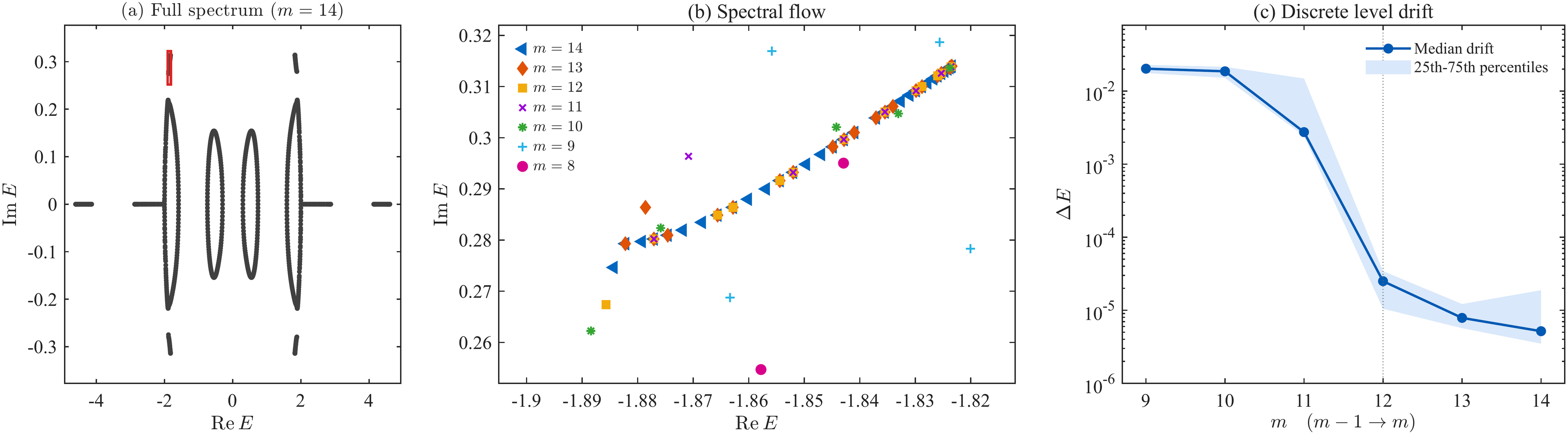}
\caption{{Size-dependent spectral evolution at the parameters of Fig.~3 of the main text, with fixed $\alpha=(\sqrt{5}-1)/2$. (a) Complete complex-energy spectrum at $m=14$ ($N=610$, $L=1220$). The red rectangle marks the upper negative-real Type-III sector enlarged in panel (b). (b) Upper negative-real type-III spectral sector for $m=8,\ldots,14$, with $L=2F_m$. (c) Median discrete level drift, defined here as the spectral-point matching distance between successive sizes [Eq.~\eqref{eq:S4drift}], obtained by one-to-one pairing that minimizes the total complex-energy distance. The blue line is the median of the matched distances; the shaded region spans their 25th--75th percentiles, calculated by linear interpolation of the sorted distances at each size comparison. It shows the spread across matched levels, not numerical error bars. The horizontal coordinate $m$ denotes the comparison $m-1\to m$; the dotted line marks the main-text crossover estimate $m\simeq12$.}}
\label{fig:spectralflow}
\end{figure}

{We examine the complex spectrum at the parameters of Fig.~3 of the main text: $t_0=1$, $\lambda=2$, $h=g=0.75$, $\theta=0$, and $\kappa=2$, under periodic boundary conditions. The modulation is fixed at $\alpha=(\sqrt{5}-1)/2$. {We use $L=2F_m$ with $m=8,\ldots,14$ and $F_0=F_1=1$, so that $m=14$ corresponds to $L=1220$. Figure~\ref{fig:spectralflow}(a) shows the complete spectrum at $m=14$ ($N=610$, $L=1220$). The red rectangle marks the upper negative-real Type-III sector enlarged in panel (b), which is selected by $\operatorname{Re}E<0$ and $\operatorname{Im}E>0.25t_0$ for the parameters considered here. As the system size increases, the spectra progressively converge toward a common spectral locus.}}

{Here, we define the \emph{discrete level drift} as the \emph{spectral-point matching distance between successive system sizes}. In plain terms, we superimpose the spectra at $m-1$ and $m$ and measure how far the paired points are apart in the complex-energy plane. A pairing rule is needed because the number of levels increases with size: the same position in a sorted list need not identify a corresponding level. We assign each point of the smaller spectral sector to a distinct point of the larger sector, choosing the pairing with the smallest sum of distances. No two old points share the same new point, and additional points of the larger sector remain unmatched. For the resulting assignment $\pi_m$, the distance for each pair is
\begin{equation}
\Delta E_n(m)=\left|E_{\pi_m(n)}^{(m)}-E_n^{(m-1)}\right|.
\label{eq:S4drift}
\end{equation}
The absolute value is the ordinary distance in the complex plane, combining changes in both the real and imaginary parts of the energy. }

{Figure~\ref{fig:spectralflow}(c) summarizes these distances for the same sector as panel (b). The line shows the median of $\Delta E_n$, representing the typical matching distance; the shaded region contains the middle $50\%$ of the distances, rather than an uncertainty in their numerical evaluation. This construction quantifies how closely successive spectra match. It does not establish that paired points represent the same physical eigenstate, and denser spectral sampling at larger sizes can also reduce the matching distances. We therefore use it as a complementary spectral diagnostic, rather than a standalone proof of localization.}

{The median drift decreases from $2.75\times10^{-3}$ for $m=10\to11$ to $2.50\times10^{-5}$ for $m=11\to12$, and reaches $5.17\times10^{-6}$ for $m=13\to14$. This pronounced suppression near $m\simeq12$ is consistent with the finite-size crossover inferred from the real-space diagnostics. Residual spectral changes remain, so these results indicate a suppression of typical level drift rather than simultaneous exact freezing of all eigenenergies.}

\section*{S5. Experimental realization}

\begin{figure}[htb]
\begin{center}
\includegraphics[width=16.8cm]{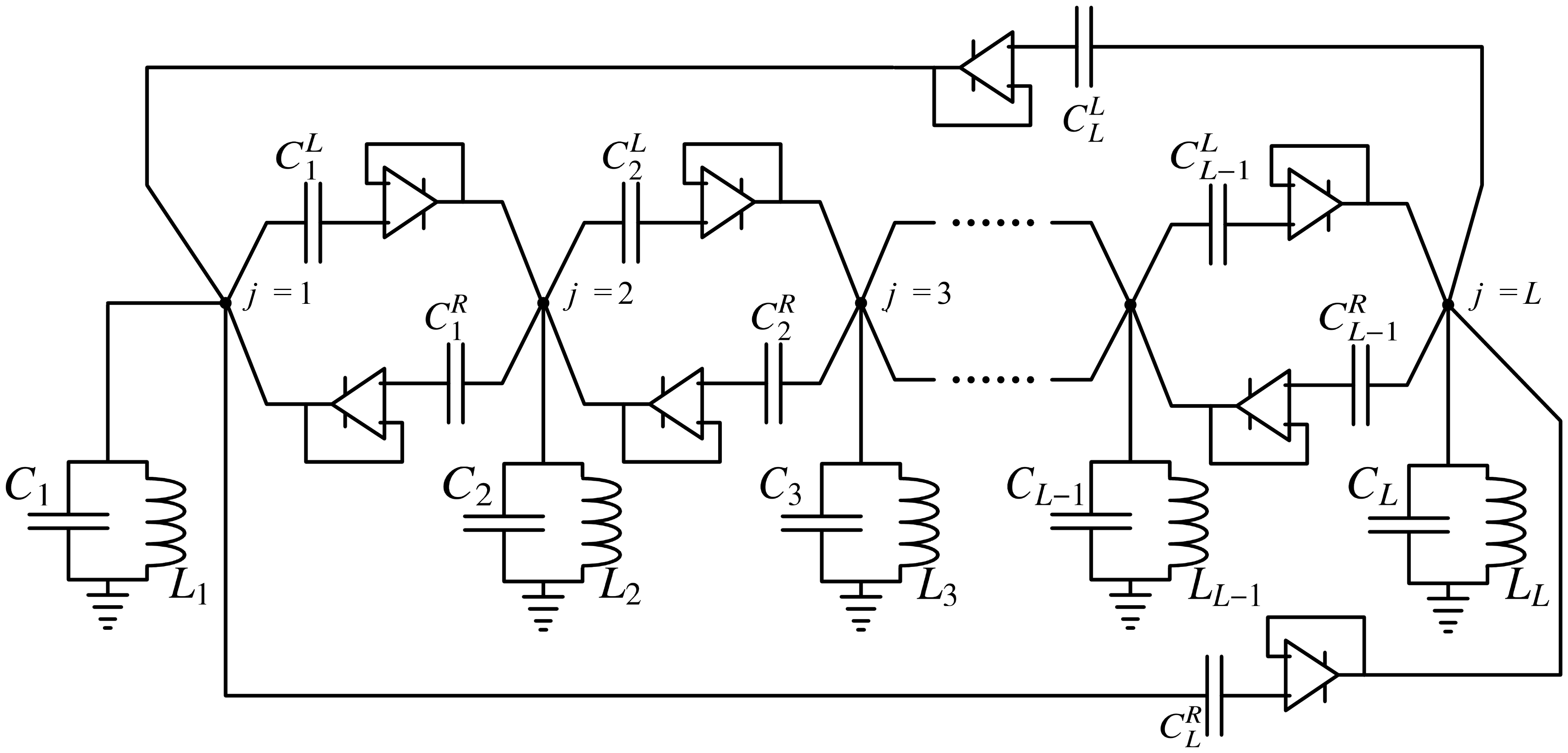}
\caption{Schematic of the classical electric circuit implementing the model. Each site $j$ contains a capacitor $C_j$ and an inductor $L_j$. Coupling capacitors $C_j^L$ and $C_j^R$, together with operational amplifiers, implement asymmetric nearest-neighbor couplings. The upper and lower connections close the chain into a periodic circuit.}
\label{fig:exp}
\end{center}
\end{figure}

In this section, we propose an experimental scheme. We design a classical circuit to realize the unidirectional quasiperiodic model, which will be instrumental in studying the pseudo-critical localization. The schematic diagram of the classical electric circuit is shown in Fig.~\ref{fig:exp}. In this circuit, we can divide the whole circuit into different unit cells. Each unit cell contains a LC resonator with a capacitor $C_j$ and inductor $L_j$, two connected capacitor $C_j^L$ and $C_j^R$, and operational amplifiers. The circuit is governed by Kirchhoff’s law $\textbf{I} = J \textbf{V}$, where $\bf{I}$ is a vector containing the current at all the sites and $\bf{V}$ is a vector containing the voltage at all the sites. The matrix $J$ is the admittance matrix at AC frequency $\omega$, which has a structure similar to the Hamiltonian, differing only by some constants. To obtain the admittance matrix $J$, we can measure the impedance matrix $G$ which is the inverse of the admittance matrix. One can see that the voltage-response measurements are performed: $\textbf{V} = G \textbf{I}$, where $\textbf{I}$ is an input current at site $j$ and the voltage is measured at different sites. Consequently, the properties of the Hamiltonian can be determined from the eigenvalues and eigenstates of the admittance matrix. The relationship between the current and voltage in the bulk can be written as
\begin{equation}
I_j = i\omega C_jV_j  + \frac{V_j}{i\omega L_j}-i\omega C_j^L\left(V_{j-1}-V_{j}\right) -i\omega C_{j+1}^R(V_{j+1}-V_{j}).
\end{equation}
We choose $C_j$ and $C_n^{L(R)}$ as the quasiperiodic potential and quasiperiodic hopping, respectively. Furthermore, we choose the inductor $L_j$ to offset the differing $i\omega(C_j^L+C_{j+1}^R)$. The admittance matrix $J$ is similar to the Hamiltonian $H$.

The current-voltage relation in the bulk can be rewritten as
\begin{equation}
I_j=
\Bigl[i\omega C_j+\frac{1}{i\omega L_j}+i\omega(C_j^L+C_{j+1}^R)\Bigr]V_j
-i\omega C_j^L V_{j-1}
-i\omega C_{j+1}^R V_{j+1}.
\end{equation}
Therefore, the admittance matrix $J$ has matrix elements
\begin{equation}
J_{j,j}=i\omega C_j+\frac{1}{i\omega L_j}+i\omega(C_j^L+C_{j+1}^R),
\end{equation}
\begin{equation}
J_{j,j-1}=-i\omega C_j^L,\qquad
J_{j,j+1}=-i\omega C_{j+1}^R,
\end{equation}
with periodic boundary terms \(J_{1,L}=-i\omega C_1^L\) and \(J_{L,1}=-i\omega C_1^R\). Equivalently, introducing the reduced admittance matrix
\begin{equation}
\tilde J \equiv \frac{J}{i\omega},
\end{equation}
one obtains
\begin{equation}
\tilde J_{j,j}=C_j-\frac{1}{\omega^2L_j}+C_j^L+C_{j+1}^R,
\end{equation}
\begin{equation}
\tilde J_{j,j-1}=-C_j^L,\qquad
\tilde J_{j,j+1}=-C_{j+1}^R.
\end{equation}
Thus \(\tilde J\) has exactly the same tridiagonal asymmetric structure as the Hamiltonian of our model, that is,
\begin{equation}
H_{j,j}=\tilde V_j,\qquad H_{j,j+1}=t_j^L,\qquad H_{j+1,j}=t_j^R,
\end{equation}
provided that we choose
\begin{equation}
C_{j+1}^R=-t_j^L,\qquad
C_{j+1}^L=-t_j^R,
\end{equation}
and
\begin{equation}
C_j-\frac{1}{\omega^2L_j}+C_j^L+C_{j+1}^R=\tilde V_j.
\end{equation} 
Here, to avoid confusion with the voltage parameter $V$, we denote the on-site potential in the model Hamiltonian as $\tilde V_j$.
In this way, the reduced admittance matrix satisfies
\begin{equation}
\tilde J=\begin{pmatrix}
\tilde\varepsilon_1 & -C_2^R & 0 & \cdots & 0 & -C_1^L \\
-C_2^L & \tilde\varepsilon_2 & -C_3^R & \cdots & 0 & 0 \\
0 & -C_3^L & \tilde\varepsilon_3 & \ddots & \vdots & \vdots \\
\vdots & \vdots & \ddots & \ddots & -C_{L-1}^R & 0 \\
0 & 0 & \cdots & -C_{L-1}^L & \tilde\varepsilon_{L-1} & -C_L^R \\
-C_1^R & 0 & \cdots & 0 & -C_L^L & \tilde\varepsilon_L
\end{pmatrix}=H.
\end{equation}
where $\tilde\varepsilon_j=C_j-\frac{1}{\omega^2L_j}+C_j^L+C_{j+1}^R$.
Negative effective couplings required in the above mapping can be implemented by active circuit elements such as operational amplifiers.

In the above, we give an experimental scheme in classical electric circuits. Here, we note that chirality can induce quantum non-reciprocity~\cite{zhang2025chirality} which potentially induces non-reciprocal dipole-dipole interaction. Therefore, based on the existing experimental scheme~\cite{zhou2023exact}, incommensurate Raman superarray of Rydberg atoms, and the chirality-induced quantum non-reciprocity, one may achieve the model Hamiltonian in the main text.

\section*{S6. Wave packet dynamics}

\begin{figure}[htb]
\begin{center}
\includegraphics[width=16.8cm]{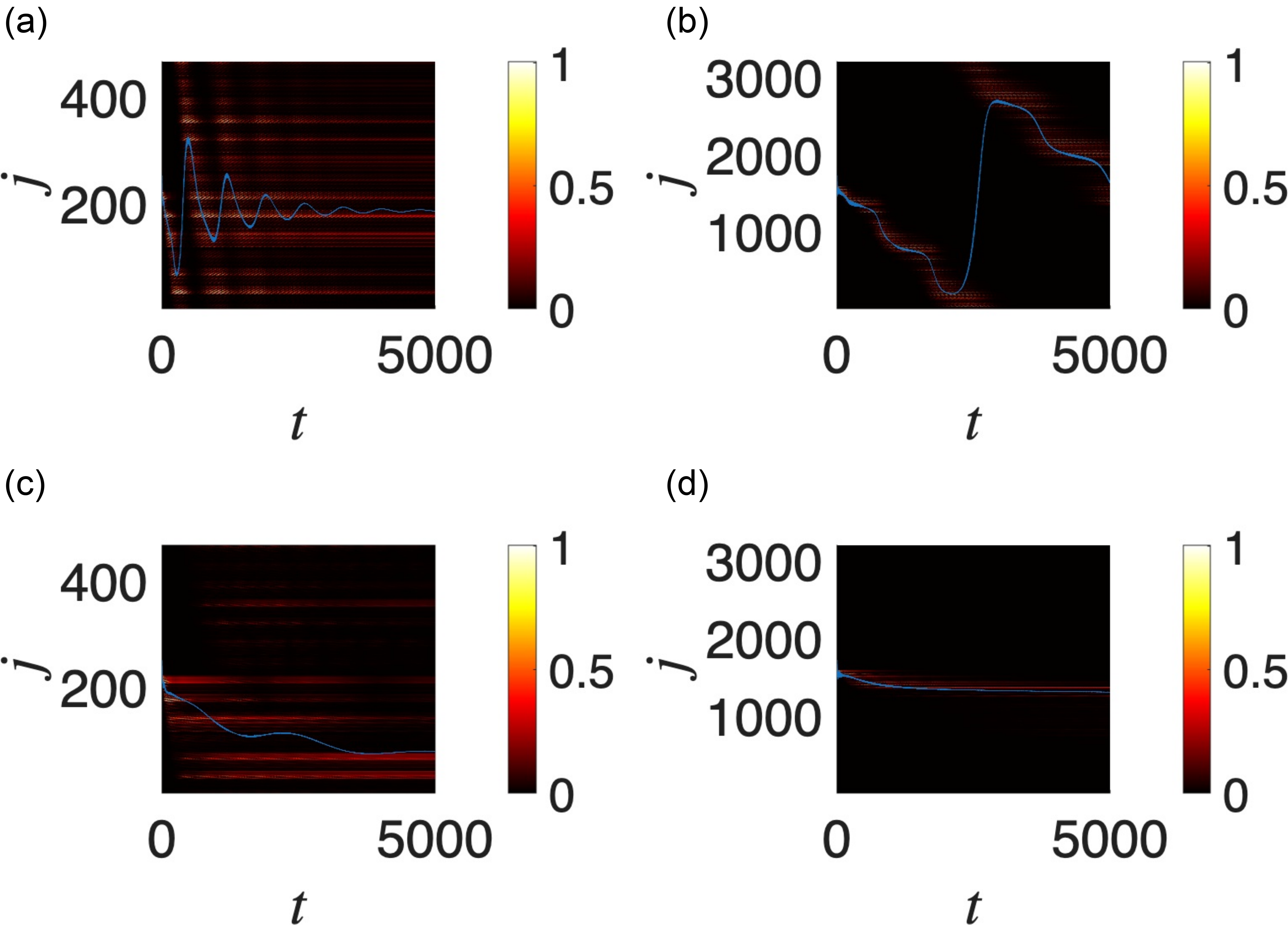}
\caption{Time evolution of the wave-function under PBCs for the initial states $\psi_j(0)=\delta_{j,N}$, with $L=2N$. In all panels, the color scale represents the normalized probability density $|\psi_j(t)|^2/C$ where $C$ is the normalization coefficient, while the blue curve shows the mean position $\langle x(t)\rangle=\sum_j j\,|\psi_j(t)|^2$.
(a,b) Dynamics for a parameter regime without pseudo-critical states, where the corresponding parameters are $h=0.7$ and $g=0.85$. The wave packet exhibits a pronounced rightward drift at early times and subsequently develops several preferred accumulation regions determined by the underlying quasiperiodic structure. Correspondingly, \(\langle x(t)\rangle\) shows pronounced oscillations before approaching a nearly stationary value. Comparing (a) with (b), the oscillation timescale becomes longer as the system size increases.
(c,d) Dynamics for a parameter regime where pseudo-critical states are present, where the corresponding parameters are $h=0.7$ and $g=0.73$. The wave packet still shows an initial rightward drift induced by the nonreciprocal hopping, but in larger systems this directed motion is restricted to an initial transient regime, after which the wave packet settles into a confined spatial region. This behavior is consistent with the finite-size crossover associated with pseudo-critical localization: at accessible sizes the dynamics retains critical-like features, while for larger systems it shows a clearer tendency toward localization. The other parameters are (a,c) $N=233$ ($L=466$), (b,d) $N=1597$ ($L=3194$), $\kappa = 2$ and $\lambda=2$.}
\label{fig:dynamics}
\end{center}
\end{figure}

In this section, we give the results of the dynamics of a wave packet initialized at a single site $\psi_j(0)=\delta_{j,j_0}$ with $j_0=N=L/2$. Figure~\ref{fig:dynamics} shows results under PBCs.

Figures~\ref{fig:dynamics}(a) and (b) correspond to a parameter regime in which no pseudo-critical states are present. In this case, the wave packet exhibits a pronounced rightward drift at early times and subsequently develops several preferred accumulation regions, reflecting the underlying quasiperiodic spatial structure. Correspondingly, the mean position $\langle x(t)\rangle=\sum_j j |\psi_j(t)|^2 $
shows clear oscillations before gradually approaching a nearly stationary value. Comparing Figs.~\ref{fig:dynamics}(a) with (b), one finds that the characteristic oscillation timescale becomes longer as the system size increases.

By contrast, Figs.~\ref{fig:dynamics}(c) and (d) correspond to a parameter regime in which pseudo-critical states are present. The wave packet still displays an initial rightward drift induced by the nonreciprocal hopping, but in larger systems this directed motion is increasingly restricted to an initial transient regime. At longer times, the wave packet no longer continues to propagate across the system, but instead settles into a confined spatial region. This behavior is compatible with the finite-size crossover associated with pseudo-critical localization: within accessible sizes the dynamics retains critical-like features, whereas for larger systems it shows a clearer tendency toward localization.

\clearpage
\section*{S7. Pseudo-critical states in non-reciprocal Aubry--Andr\'e model}
In this section, we study the pseudo-critical states in an non-reciprocal Aubry--Andr\'e model. This model features a coexistence of delocalized and Anderson localized states, and its Hamiltonian is given by~\cite{li2024asymmetric}:
\begin{equation}
H = \sum_{j=1}^{L-1} \left((t+W_j)\hat{a}_j^\dagger\hat{a}_{j+1} + t\hat{a}_{j+1}^\dagger\hat{a}_j\right) + \sum_{j=1}^{L}V_j\hat{a}_j^\dagger\hat{a}_{j},
\end{equation}
where 
\begin{equation}
\{W_j,V_j\} = \left\lbrace\begin{matrix}
\{W,2\lambda\}\cos(2\pi\alpha j+\theta),	&j=n\kappa,\\
0,&\text{else},\\
\end{matrix}\right.
\end{equation}
with $\kappa$ ($\kappa \leq 2$) an integer, $\lambda$ the hopping coefficient, and $\theta$ the phase offset. Here, to induce pseudo-critical states, we introduce an non-reciprocal hopping term
\begin{equation}
H_g = \sum_{j\neq n\kappa}^{L-1}\left(g\hat{a}_j^\dagger\hat{a}_{j+1}-g\hat{a}_{j+1}^\dagger\hat{a}_j\right).
\end{equation}
Then the total Hamiltonian is $H+H_g$. The transfer matrices can be obtained as follows:
\begin{equation}
\begin{aligned}
\mathcal{T}_i =& \begin{pmatrix}
\frac{E}{1+g} 	&\frac{-(1-g)}{1+g}\\
1 &0
\end{pmatrix}^{\kappa-2}
\begin{pmatrix}
\frac{E}{1+g}	&\frac{-1}{1+g}\\
1	&0
\end{pmatrix}
\begin{pmatrix}
\frac{E-V_j}{1+W_j}	&\frac{-(1-g)}{1+W_j}\\
1	&0
\end{pmatrix}\\
=&\frac{1}{1+W_j}\begin{pmatrix}
a_\kappa	&-\alpha a_{\kappa-1}\\
a_{\kappa-1} &-\alpha a_{\kappa-2}\\
\end{pmatrix}
\begin{pmatrix}
\frac{E^2-EV_j-W_j-1}{1+g}	&\frac{-E(1-g)}{1+g}\\
E-V_j	&-(1-g)\\
\end{pmatrix},
\end{aligned}
\end{equation}
where 
\begin{equation}
 \begin{pmatrix}
\frac{E}{1+g} 	&\frac{-(1-g)}{1+g}\\
1 &0
\end{pmatrix}^{\kappa-2}
=
\begin{pmatrix}
a_\kappa	&-\alpha a_{\kappa-1}\\
a_{\kappa-1} &-\alpha a_{\kappa-2}\\
\end{pmatrix},
\end{equation}
and $a_\kappa$ is defined as
\begin{equation}
a_\kappa = \frac{1}{\Delta}\left[\left(\frac{\beta+\Delta}{2}\right)^{\kappa-1}
-\left(\frac{\beta-\Delta}{2}\right)^{\kappa-1}\right],
\end{equation}
with $\Delta = \sqrt{\beta^2-4\alpha}$, $\alpha = (1-g)/(1+g)$ and $\beta=E/(1+g)$. Similar to the previous approach, we complexify the phase via the substitution $\theta\rightarrow \theta+i\epsilon$ and combine this transformation with Avila’s global theory to obtain 
\begin{equation}
\gamma(E) = \left\{
\begin{matrix}
\max\left\{\frac{1}{\kappa}\ln\left|\frac{2K_\kappa}{1+\sqrt{1-W^2}}\right|,0\right\}, &1\geq W,\\
\max\left\{\frac{1}{\kappa}\ln\left|\frac{2K_\kappa}{W}\right|,0\right\},	&1<W,
\end{matrix}
\right.
\end{equation}
where $K_\kappa = -A a_\kappa+\alpha\lambda a_{\kappa-1}$ with $A=\frac{E\lambda+\frac{W}{2}}{1+g}$. When $\kappa = 2$ and $W>1$, one can get that the mobility ring is determined by
\begin{equation}
\frac{1}{2}\ln\left|\frac{2E\lambda+W}{W(1+g)}\right| = 0.
\label{eqS:mr}
\end{equation}

\begin{figure}[H]
\begin{center}
\includegraphics[width=16.8cm]{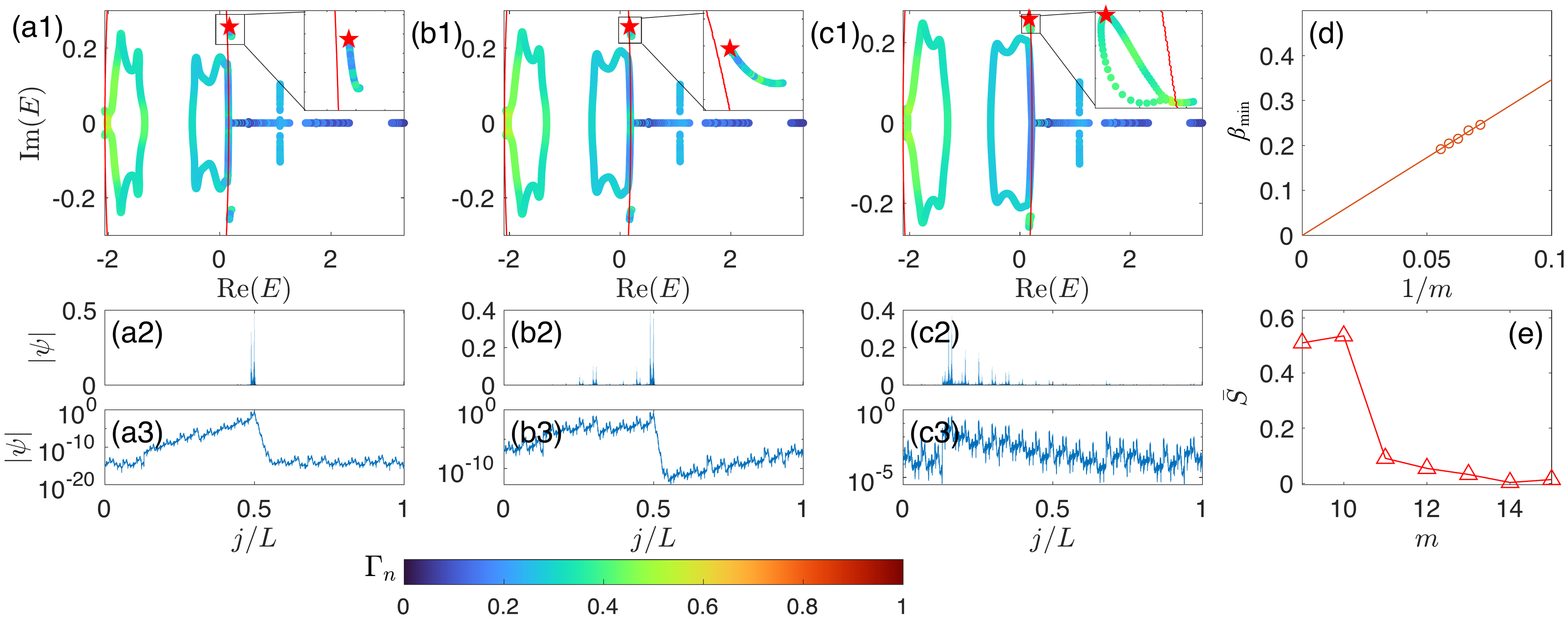}
\caption{
(a1) to (c1) show the eigenenergies marked by the fractal dimension $\Gamma_n$ for the corresponding eigenstates
for $g=(0.17,0.2,0.23)$, respectively.
The insets in (a1)--(c1) enlarge the upper complex-energy sector containing the starred eigenstates. The red curves represent the mobility rings obtained from Eq.~\eqref{eqS:mr}. 
(a2) and (a3) show the wavefunctions of the local skin states corresponding to the red star in (a1). (b2) and (b3) show the wavefunctions of the pseudo-critical states corresponding to the red star in (b1). (c2) and (c3) show the wavefunctions of the critical states corresponding to the red star in (c1). (d) and (e) present the multifractal analysis and the average entanglement entropy, respectively, of the pseudo-critical states for the parameter set used in (b). Other parameters are $W=1.9$, $\lambda=1$, $t=1$ and $N=1597$.
}
\label{fig:irrecipAA}
\end{center}
\end{figure}

Figs.~\ref{fig:irrecipAA}(a) to (c) display the eigenenergies marked by the fractal dimension $\Gamma_n$ for the corresponding eigenstates

for $g=(0.17,0.2,0.23)$, respectively,
and a transition from localized states to critical states (with fractal structures) is found for the eigenstates in the enlarged insets [also for those with symmetric negative ${\rm Im}(E)$].
In contrast to the model in the main text,
here these states show pseudo-critical behaviors when their eigenenergies lie outside—yet in close proximity to—the mobility ring [those in panel (b)], as shown by the average scaling exponent $\beta_{\text{min}}$ and entanglement entropy $\bar{S}$ in Figs.~\ref{fig:irrecipAA}(d) and (e), respectively.
In Fig.~\ref{fig:irrecipAA}(c), the concerned eigenstates possess a loop-like spectrum and become critical states.

\par

\bibliographystyle{scpma}
\bibliography{refs}